\documentclass[aapm,mph,preprint,amsmath,amssymb]{revtex4-1}

\usepackage[mathlines]{lineno}
\modulolinenumbers[5]
\linenumbers\relax 

\pdfoutput=1
\usepackage[pdftex]{graphicx}
\usepackage{algorithm}
\usepackage{algpseudocode}
\usepackage{caption}

\newcommand{\iv}[1]{\mathbf{#1}}
\newcommand{\sm}[1]{\mathit{#1}}

\DeclareMathOperator*{\argmax}{arg\,max}
\DeclareMathOperator*{\argmin}{arg\,min}

\begin{document}
\captionsetup[algorithm]{font=small,labelfont=bf,singlelinecheck=false,justification=raggedright,labelsep=colon}
\captionsetup[figure]{font=small,labelfont=bf,singlelinecheck=false,justification=raggedright,labelsep=colon}

\title{First-order convex feasibility algorithms for 
X-ray CT
}

\author{Emil Y. Sidky}
\email{sidky@uchicago.edu}
\affiliation{%
University of Chicago \\
Department of Radiology \\
5841 S. Maryland Ave., Chicago IL, 60637
}%

\author{Jakob S. J{\o}rgensen}
\email{jakj@imm.dtu.dk}
\affiliation{%
Technical University of Denmark\\
Department of Informatics and Mathematical Modeling\\
Richard Petersens Plads, Building 321, 2800 Kgs. Lyngby, Denmark
}%

\author{Xiaochuan Pan}
\email{xpan@uchicago.edu}
\affiliation{%
University of Chicago \\
Department of Radiology \\
5841 S. Maryland Ave., Chicago IL, 60637
}%

\date{\today}

\begin{abstract}
PURPOSE: Iterative image reconstruction (IIR) algorithms in Computed Tomography (CT) are based
on algorithms for solving a particular optimization problem.  Design of the IIR algorithm, therefore,
is aided by knowledge of the solution to the optimization problem on which it is based.
Often times, however, it is impractical to
achieve accurate solution to the optimization problem of interest, which complicates design of
IIR algorithms. This issue is particularly acute for CT with a limited angular-range scan,
which leads to poorly conditioned system matrices and difficult to solve optimization
problems.  In this article, we develop IIR algorithms which solve a certain type of
optimization called convex feasibility. The convex feasibility approach can provide
alternatives to unconstrained
optimization approaches and at the same time allow for rapidly convergent
algorithms for their solution -- thereby facilitating the IIR algorithm design process.\\
\noindent

METHOD: An accelerated version of the Chambolle-Pock (CP) algorithm is adapted to
various convex feasibility problems of potential interest to IIR in CT. One of the
proposed problems is seen to be equivalent to least-squares minimization, and two
other problems provide alternatives to
penalized, least-squares minimization.\\
\noindent

RESULTS: The accelerated CP algorithms are demonstrated on a simulation of circular fan-beam
CT with a limited scanning arc of 144$^\circ$. The CP algorithms are seen in
the empirical results to converge to the solution of
their respective convex feasibility problems.\\
\noindent

CONCLUSION: Formulation of convex feasibility problems can provide
a useful alternative to unconstrained 
optimization when designing IIR algorithms for CT. The approach
is amenable to recent methods for accelerating first-order
algorithms which may be particularly useful for CT with limited
angular-range scanning. The present article demonstrates the methodology, and future
work will illustrate its utility in actual CT application.\\
\noindent
\end{abstract}


\maketitle

\section{Introduction}

Iterative image reconstruction (IIR) algorithms in computed tomography (CT) are designed based
on some form of optimization.  When designing IIR algorithms
to account for various factors in the CT model, the actual designing occurs usually
at the optimization problem and not the individual processing steps of the IIR algorithm. Once the
optimization problem is established, algorithms are developed to solve it. 
Achieving convergent algorithms is important, because
they yield access to the designed solution of the optimization problem and allow for
direct assessment
of what factors to include in a particular optimization problem.
Convergent algorithms can also aid in determining at what iteration number to truncate an IIR
algorithm.
With access to the designed solution,
the difference between it and previous iterates can be quantitatively
evaluated to see whether this difference
is significant with respect to a given CT imaging task.

It can be challenging to develop convergent algorithms for some optimization problems
of interest. This issue is particularly acute for CT,
which involves large-scale optimization. In using the term ``large-scale'', we are specifically
referring to optimization problems based on a linear data model, and the dimension of the linear system
is so large that the system matrix cannot be explicitly computed and stored in memory.
Such systems only allow for algorithms which employ operations of a similar
computational expense to matrix-vector products.
Large-scale optimization algorithms are generally restricted to first-order methods,
where only gradient information on the objective function is used, or row-action algorithms such
as the algebraic reconstruction technique (ART) \cite{Gordon:70,Herman}. Recently,
there has been renewed interest in developing convergent algorithms for
optimization problems involving $\ell_1$-based
image norms, and only in the last couple of years have practical, convergent algorithms
been developed to solve these optimization problems for IIR in CT
\cite{Jensen2011,Defrise:11,fessler:2012,SidkyCP:2012}. Despite the progress in algorithms,
there
are still CT configurations of practical interest, which can lead to optimization
problems that can be quite challenging to solve accurately. Of particular interest
in this work is CT with a limited angular-range scanning arc. Such a configuration 
is relevant to many C-arm CT and tomosynthesis applications.
Modeling limited angular-range scanning, 
leads to system matrices with unfavorable singular value
spectra and optimization problems for which many algorithms converge slowly.

In this article, we consider application
of convex feasibility \cite{combettes1996convex,combettes1993foundations} to IIR for
CT. In convex feasibility, various constraints
on properties of the image are formulated so that each of these constraints specifies a convex set.
Taking the intersection of all of the convex sets yields a single convex set, and the idea is to
simply choose one of these images in the intersection set.
We have found convex feasibility to be useful for CT IIR algorithm design \cite{Han:2012},
and it is of particular interest here for limited angular-range
CT, because convex feasibility is amenable to
recent accelerated first-order algorithms proposed by Chambolle and Pock (CP) \cite{chambolle2011first}.
In Sec. \ref{sec:methods}, we specify the limited angular-range CT system, discuss
unconstrained optimization approaches, and then list three useful convex feasibility
problems along with a corresponding accelerated CP algorithm.
In Sec. \ref{sec:results}, the accelerated convex feasibility CP algorithms are
demonstrated with simulated CT projection data.

\section{Methods: Chambolle-Pock algorithms for convex feasibility}
\label{sec:methods}

For this article, we focus on modeling circular, fan-beam CT with a limited scanning angular range.
As with most work on IIR, the data model is discrete-to-discrete (DD) and can be written
as a linear equation
\begin{equation}
\label{xfeg}
\iv{g} = \sm{X} \iv{f},
\end{equation}
where $\iv{f}$ is the image vector comprised of pixel coefficients, $\sm{X}$ is the system
matrix generated by computing the ray-integrals with the line-intersection method, and $\iv{g}$
is the data vector containing the estimated projection samples.
For the present investigation on IIR algorithms, we consider a
single configuration for limited angular range scanning where the system matrix $\sm{X}$
has a left-inverse ($\sm{X}^T \sm{X}$ is invertible) but is numerically unstable in the sense that it has
a large condition number. The vector $\iv{f}$ consists of the pixels within a
circle inscribed in a 256$\times$256 pixel array; the total number of pixels is
51,468.  The sinogram contains 128 views spanning a 144$^\circ$ scanning arc, and
the projections are taken on a 512-bin linear detector array. The modeled source-to-isocenter
and source-to-detector distances are 40 and 80 cm, respectively.  The total number
of transmission measurements is 65,536, and as a result the system matrix $\sm{X}$
has about 25\% more rows than columns.
The condition of $\sm{X}$, however, is poor,
which can be understood by considering the corresponding continuous-to-continuous (CC)
fan-beam transform. A sufficient angular range for stable inversion of the CC fan-beam
transform requires a 208$^\circ$ scanning arc (180$^\circ$ plus the fan-angle,
see for example Sec. 3.5 of Ref. \cite{Kak}).
By using the inverse power method, as described in Ref. \cite{jakob:2011}, the condition
number, the ratio of the largest to smallest singular value,
for $\sm{X}$ is determined to be $2.55 \times 10^{4}$.
One effect of the large
condition number is to amplify noise present in the data, but it also can cause
slow convergence for optimization-based IIR.

\subsection{Unconstrained optimization for IIR in CT}
Image reconstruction using this DD data model is usually performed with some
form of optimization, because physical factors and inaccuracy of the model render
Eq. (\ref{xfeg}) inconsistent -- namely, no $\iv{f}$ exists satisfying this equation.
Typically in using this model, quadratic optimization problems
are formulated, the simplest of
which is the least-squares problem
\begin{equation}
\label{lsq}
\iv{f}^\circ = \argmin_\iv{f} \left\{ \frac{1}{2} \| \iv{g} - \sm{X} \iv{f} \|^2_2 \right\},
\end{equation}
where $\iv{f}^\circ$ is the image which minimizes the Euclidean distance
between the available data $\iv{g}$ and the estimated data $\sm{X} \iv{f}$.
In the remainder of the article, we use the superscript ``$^\circ$'' to indicate
a solution to an optimization problem.
Taking the gradient of this objective function, and setting it to zero component-wise,
leads to the following consistent linear equation
\begin{equation}
\label{xfegConsistent}
\sm{X}^T \sm{X} \iv{f}=\sm{X}^T \iv{g},
\end{equation}
where the superscript $T$ denotes the matrix transpose. This linear equation
is particularly useful for setting up
the linear conjugate gradients (CG) algorithm, see for example
Ref. \cite{Nocedal:06}, which has been used as the gold standard algorithm
for large-scale
quadratic optimization problems in IIR. The reader is also referred to conjugate gradients
least-squares (CGLS) and LSQR (an algorithm for sparse linear
equations and sparse least squares),
which solve Eq. (\ref{lsq}) for non-symmetric $\sm{X}$ \cite{Paige:1982}.


The solution to Eq. (\ref{lsq}) or (\ref{xfegConsistent}) can be undesirable because
of inconsistency in the data. Particularly for the present case, the poor conditioning of $\sm{X}$
can yield tremendously amplified artifacts in the reconstructed image. As is well-known, artifacts
due to data inconsistency can be controlled in optimization-based IIR by adding a penalty term to
discourage large variations between neighboring pixels
\begin{equation}
\label{reglsq}
\iv{f}^\circ = \argmin_\iv{f} \left\{ \frac{1}{2} \| \iv{g} - \sm{X} \iv{f} \|^2_2 +
\alpha R(\iv{f}) \right\},
\end{equation}
where $R(\iv{f})$ is a generic roughness term which usually is a convex function of
the difference between neighboring pixels in the image. The parameter $\alpha$ controls
the strength of the penalty with larger values leading to smoother images. When $R(\iv{f})$
is chosen to be quadratic in the pixel values, the optimization problem can be solved
by a host of standard algorithms including CG. Of recent interest have been convex regularizers
based on the $\ell_1$-norm, which is more difficult to treat and, accordingly, for which
many new, convergent algorithms have been proposed and applied to image reconstruction
in CT \cite{Jensen2011,Defrise:11,fessler:2012,SidkyCP:2012}.

\subsection{Convex feasibility}
In this article, we consider convex feasibility problems which
provide alternatives to the above-mentioned optimization problems.
For convex feasibility problems, 
convex sets resulting from constraints on various properties
of the image are formulated, and a single image which satisfies all the imposed
constraints is sought. 
Most algorithms for such problems are 
based on projection onto convex sets (POCS)
\cite{combettes1993foundations}, where the image estimate is sequentially projected onto each
constraint set. Convex feasibility problems can be: inconsistent, no image satisfies all the constraints;
or consistent, at least one image satisfies all the constraints. In either case, 
POCS algorithms can yield a useful solution. In the inconsistent case, POCS algorithms can
be designed to yield an image ``close'' to satisfying all the constraints. In the consistent case, 
a POCS algorithm can be designed to find an image obeying all the constraints. In either case,
the issue of uniqueness is secondary, as an image ``in the middle'' of many inconsistent constraints
or in the intersection set of consistent constraints is considered to be equally valid.
Accordingly, the POCS result often depends on starting image, relaxation schemes, and projection order.

For our purposes we write a general convex feasibility
as the
following optimization problem
\begin{equation}
\label{cfgen0}
\iv{f}^\circ = \argmin_\iv{f} \left\{  
\sum_i \delta_{S_i}(\sm{K}_i (\iv{f}))  \right\},
\end{equation}
$\sm{K}_i(\cdot)$ is the $i$th affine transform of the image $\iv{f}$;
$S_i$ is the $i$th convex set to which $\sm{K}_i (\iv{f})$ belongs;
and the indicator function $\delta$ is defined
\begin{equation}
\label{indicator}
\delta_S(\iv{x}) =
\begin{cases}
0 & \iv{x} \in S \\
\infty & \iv{x} \not\in S
\end{cases}.
\end{equation}
The use of indicator functions in convex analysis provides a means to turn convex
sets into convex functions \cite{rockafellar1970convex}, and in this case, they allow
convex feasibility problems
to be written as a minimization of a single objective function.
The objective function in Eq. (\ref{cfgen0}) is zero for any image $\iv{f}$ satisfying all the constraints,
i.e. $\sm{K}_i (\iv{f}) \in S_i$ for all $i$, and it is infinity if any of the constraints
are violated. For a consistent convex feasibility problem,
the objective minimum is zero, and for an inconsistent 
convex feasibility problem, the objective minimum is infinity.

\subsection{Modified convex feasibility optimization and the Chambolle-Pock
primal-dual algorithm }
To solve the generic convex feasibility problem in Eq. (\ref{cfgen0}), we modify this optimization
problem by adding a quadratic term
\begin{equation}
\label{cfgen}
\iv{f}^\circ = \argmin_\iv{f} \left\{ \frac{1}{2} \| \iv{f} - \iv{f}_\text{prior} \|^2_2
+ \sum_i \delta_{S_i}(\sm{K}_i (\iv{f}))  \right\},
\end{equation}
where $\iv{f}_\text{prior}$ is a prior image estimate that
can be set to zero if no prior image is available.
With this optimization problem, we actually specify a unique solution to our generic
convex feasibility problem in
the consistent case -- namely the image satisfying all constraints and closest to
$\iv{f}_\text{prior}$. As we will demonstrate the algorithm we propose to use for solving
Eq. (\ref{cfgen}) appears to yield useful solutions for the {\it inconsistent} case.
This latter property can be important for
IIR in CT because the data model in Eq. (\ref{xfeg}) is often inconsistent with the available
projection data.

The reason for recasting the optimization problem in the form shown in Eq. (\ref{cfgen}) is that
this optimization problem can be solved by an accelerated algorithm described in Ref.~\cite{chambolle2011first}.
Recently, we have been interested in a convex optimization framework
and algorithms derived by Chambolle and Pock (CP) \cite{chambolle2011first,Pock2011}.
This framework centers on the generic convex optimization problem
\begin{equation}
\label{primal}
p^\circ=\min_{\iv{x}} \; \left\{ G(\iv{x}) + F(\sm{H} \iv{x}) \right\},
\end{equation}
where $G(\cdot)$ and $F(\cdot)$ are convex functions, and $\sm{H}$ is a linear transform.
The objective function
\begin{equation}
\notag
p= G(\iv{x}) + F(\sm{H} \iv{x})
\end{equation}
is referred to as the primal objective.
This generic problem encompasses many optimization problems of interest to IIR in CT, because non-smooth
convex functions such as the indicator and $\ell_1$-norm can be incorporated into $F$ or $G$.
Also, the linear transform $\sm{H}$ can model projection, for a data fidelity term,
or a finite-difference-based
gradient, for an image total variation (TV) term. The CP framework, as presented
in Ref. \cite{chambolle2011first}, comes with four algorithms that have different worst-case
convergence rates depending on convexity properties of $F$ and $G$.
Let $N$ be the number of iterations, the algorithm summaries are:
\begin{description}
\item[CP Algorithm 1] This basic CP algorithm forms the basis of the subsequent
algorithms and it only requires $F$ and $G$ to be convex.
The worst-case convergence rate is $O(1/N)$.
\item[CP Algorithm 2] Can be used if either $F$ or $G$ are uniformly convex.
Modifies CP Algorithm 1 using a step-size formula developed by
Nesterov \cite{nesterov1983method,beck:09}. The worst-case convergence rate 
is $O(1/N^2)$. Because the convergence rate is faster than the
previous case, this algorithm is an accelerated version of CP Algorithm 1.
\item[CP Algorithm 3] Can be used if both $F$ and $G$ are uniformly convex.
This algorithm is the same as CP Algorithm 1, except that there is a specific
choice of algorithm parameters, depending on constants related to
the uniform convexity of $F$ and $G$.
The worst-case convergence is linear, i.e. $O(1/c^N)$, where $c>1$ is a constant.
\item[CP Algorithm 4] A simpler version of CP Algorithm 2, which
also requires $F$ or $G$ to be uniformly convex. The convergence rate
is slightly worse than $O(1/N^2)$.
\end{description}

In a previous publication \cite{SidkyCP:2012},
we illustrated how to use CP Algorithm 1 from Ref. \cite{chambolle2011first}
to prototype many optimization
problems of potential interest to image reconstruction in CT.
We were restricted to CP Algorithm 1, because we considered mainly
problem where $G$ was 0, and $F$ contained indicator functions,
the $\ell_1$-norm, or TV terms and accordingly $F$ was not uniformly convex.
In the present work, we narrow
the class of optimization problems to those which can be written
in the form of Eq. (\ref{cfgen}),
where the sets $S_i$ are simple enough that direct Euclidean
projections to the sets $S_i$ are analytically available.
In matching up Eq. (\ref{cfgen}) to the generic optimization problem in Eq. (\ref{primal})
the function $G$ is assigned the uniformly convex quadratic term and $F$
gets the sum of indicator functions. As such, Eq. (\ref{cfgen}) fills the
requirements of CP Algorithms 2 and 4. In the particular case of Eq. (\ref{cfgen})
the uniformly convex term, $ 0.5 \| \iv{f} - \iv{f}_\text{prior} \|^2_2$,
is simple enough that CP Algorithm 2 can be derived without any difficulty.
Because this algorithm is an accelerated version of CP Algorithm 1, we refer
to it, here, as the accelerated CP
algorithm. This algorithm acceleration
is particularly important for IIR involving an ill-conditioned
data model such as Eq. (\ref{xfeg})
in the case of limited angular range scanning.

\subsection{The primal-dual gap and convergence criteria}

The CP algorithms are primal-dual in that they solve the primal minimization problem
Eq. (\ref{primal})
together with a dual maximization problem
\begin{equation}
\label{dual}
d^\circ=\max_{\iv{y}} \; \left\{ -F^*(\iv{y}) - G^*(-\sm{H}^T \iv{y}) \right\},
\end{equation}
and,
\begin{equation}
\notag
d= -F^*(\iv{y}) - G^*(-\sm{H}^T \iv{y})
\end{equation}
is the dual objective function,
and the superscript
$^*$ represents convex conjugation through the Legendre transform
\begin{equation}
\label{legendre}
P^*(\iv{z})  = \max_{\iv{z}^\prime}  \; \left\{  \iv{z}^T \iv{z}^\prime
- P(\iv{z}^\prime) \right\}.
\end{equation}
That the CP algorithms obtain the dual solution, also, is useful for
obtaining a robust convergence criterion that applies for
non-smooth convex optimization. As long as the primal
objective function $p$ is convex, we have $p^\circ = d^\circ$.
While a solution for a smooth optimization problem
can be checked by observing that the gradient of the primal
objective function in Eq. (\ref{primal}) is zero, this test may not
be applicable to non-smooth
optimization problems, where the primal objective function may not be differentiable at
its minimum.
Instead, we can use the primal-dual gap $p-d$, because the primal objective function
for any $\iv{x}$ is larger than the dual objective function in Eq. (\ref{dual})
for any $\iv{y}$ except when $\iv{x}$ and $\iv{y}$ are at their
respective extrema, where these objective functions are equal.
Checking the primal-dual gap is complicated slightly when indicator
functions are included in one of the objective functions, because indicators
take on infinite values when their corresponding constraint is not
satisfied. As a result, we have found it convenient
in \cite{SidkyCP:2012} to define a conditional primal-dual gap
which is the primal-dual gap with indicator functions removed
from both objective functions. This convergence check then involves
observing that the conditional primal-dual gap is tending
to zero and that the iterates are tending toward satisfying
each of the constraints corresponding to the indicator functions.
By dividing up the convergence check in this way, we give up
non-negativity of the gap. The conditional primal-dual gap can
be negative, but it will approach zero as the iterates approach
the solution to their respective optimization problems. Use of this
convergence check will become more clear in the results section
where it is applied to various convex feasibility problems related
to IIR in CT.

With respect to numerical convergence, it is certainly useful to
have mathematical convergence criteria such as the gradient of the
objective function or the primal-dual gap, but it is also important to
consider metrics of interest.
By a metric, we mean some
function of the image pixel values pertaining to a particular
purpose or imaging task.
For numerical convergence, we need to
check, both, that the convergence metrics are approaching zero and 
that other metrics of interest are leveling off so that they do not
change with further iterations. Rarely
are IIR algorithms run to the point where the convergence criterion
are met exactly, in the numerical sense. This means, that the image
estimates are still evolving up until the last computed iteration,
and one cannot say {\it a priori} whether the small changes in the
image estimates are important to the metrics of interest or not.  For the
present theoretical work, where we have access to the true
underlying image, we employ the image root mean square error (RMSE)
as an image quality metric. But we point out that other metrics may be
more sensitive and potentially alter the iteration number where
the specific problem can be considered as converged \cite{Nonuniform_MIC}.

\subsection{Convex feasibility instances}
In the following, we write various imaging problems
in the form of Eq. (\ref{cfgen}). We consider the following
three convex feasibility problems: EC, one set specifying a data equality constraint;
IC, one set specifying a data inequality constraint; and ICTV, two sets specifying
data and TV inequality constraints.
The derived accelerated CP
algorithms for each problem are labeled CP2-EC, CP2-IC, and CP2-ICTV, respectively.
Using simulated fan-beam CT data with a limited angular-range scanning arc,
Sec. \ref{sec:results} presents results for all three
problems in the consistent case and
problems EC and ICTV in the inconsistent case. Of particular importance, CP2-EC
applied to the inconsistent case appears to
solve the ubiquitous least-squares optimization problem with a convergence
rate competitive with CG.


\subsubsection{CP2-EC: an accelerated CP algorithm instance for a data equality constraint}
\label{sec:firstalg}

The data model in Eq. (\ref{xfeg}) cannot be used directly as an implicit imaging model for
real CT data, because inconsistencies inherent in the data prevent a solution. But
treating this equation as an implicit imaging model for ideal simulation can be useful for
algorithm comparison and testing implementations of the system matrix $\sm{X}$; we use
it for the former purpose. We write this ideal
imaging problem into an instance of Eq. (\ref{cfgen})
\begin{equation}
\label{cfxfeg}
\iv{f}^\circ = \argmin_\iv{f} \left\{ \frac{1}{2} \| \iv{f} - \iv{f}_\text{prior} \|^2_2
+ \delta_{0}(\sm{X} \iv{f} -\iv{g})  \right\},
\end{equation}
where the indicator $\delta_0(\cdot)$ is zero only when all components of the argument vector
are zero, and otherwise it is infinity. The corresponding dual maximization problem
needed for computing the conditional primal-dual gap is
\begin{equation}
\label{cfxfegdual}
\iv{y}^\circ = \argmax_\iv{y} \left\{- \frac{1}{2} \|X^T \iv{y} \|^2_2
-\iv{g}^T \iv{y} + \iv{f}^T_\text{prior} (X^T \iv{y})
\right\}.
\end{equation}
In matching Eq. (\ref{cfxfeg}) with
Eq. (\ref{cfgen}), there is only one convex constraint where $\sm{K}_1(\iv{f}) = \sm{X} \iv{f} - \iv{g}$ and
$S_1$ is the 0-vector with size, $\text{size}(\iv{g})$.  In considering ideal data and a left-invertible
system matrix $\sm{X}$, there is only one image for which the indicator is not infinite.
In this situation, the first quadratic has no effect on the solution and accordingly
the solution is independent of the prior image estimate $\iv{f}_\text{prior}$. If the system matrix is not
left-invertible, the solution to Eq. (\ref{cfxfeg}) is
the image satisfying Eq. (\ref{xfeg}) closest to $\iv{f}_\text{prior}$.

\begin{figure}
\hrulefill
\begin{algorithmic}[1]
\State $L \gets \| \sm{X} \|_2; \; \tau \gets 1; \; \sigma \gets 1/L^2; \; n \gets 0$
\State initialize $\iv{f}_0$ and $\iv{y}_0$ to zero vectors
\State $\bar{\iv{f}}_0 \gets \iv{f}_0$
\Repeat
\State $\iv{y}_{n+1} \gets \iv{y}_n + \sigma( \sm{X} \bar{\iv{f}}_n-\iv{g})$ \label{dualupdate1}
\State $\iv{f}_{n+1} \gets \left[ \iv{f}_n - \tau (\sm{X}^T \iv{y}_{n+1} - \iv{f}_\text{prior})
\right] / (1+\tau)$ \label{primalupdate1}
\State $ \theta \gets 1/\sqrt{1+2\tau}$; $\tau \gets \tau \theta$; $\sigma \gets \sigma / \theta$ \label{fancy}
\State $\bar{\iv{f}}_{n+1} \gets \iv{f}_{n+1} + \theta(\iv{f}_{n+1} - \iv{f}_n)$
\State $n \gets n+1$
\Until{$n \ge N$}
\end{algorithmic}
\hrulefill
\caption{Pseudocode for $N$ steps of the accelerated CP algorithm instance
for solving Eq. (\ref{cfxfeg}). Variables are defined in the text.}
\label{alg:cfxfeg}
\end{figure}

Following the formalism of Ref. \cite{chambolle2011first}, we write an accelerated CP
algorithm instance 
for solving Eq. (\ref{cfxfeg}) and its dual
Eq. (\ref{cfxfegdual}) in Fig. \ref{alg:cfxfeg}.
We define the pseudocode
variables and operations starting from the first line.  The variable $L$ is assigned
the matrix $\ell_2$-norm of $\sm{X}$, which is its largest singular value. This quantity
can be computed by the standard power method, see \cite{SidkyCP:2012} for its application
in the present context.
The parameters $\tau$ and $\sigma$ control the step sizes in the primal and dual problems
respectively, and they are initialized so that their product yields $1/L^2$. Other choices
on how to balance the starting values of $\tau$ and $\sigma$ can be made, but we have
found that the convergence of our examples does not depend strongly on the choice
of these parameters. Line \ref{dualupdate1} shows the update of the dual variable $\iv{y}_{n+1}$;
this variable has the same dimension as the data vector $\iv{g}$.
Line \ref{primalupdate1} updates the image, and Line \ref{fancy} adjusts the step-sizes
in a way that accelerates the CP algorithm \cite{chambolle2011first}.

\subsubsection{CP2-IC: an accelerated CP 
algorithm instance for inequality constrained data-error}
\label{sec:secondalg}

Performing IIR with projection data containing inconsistency, requires some form
of image regularization. One common strategy is to employ Tikhonov regularization,
see for example Chapter 2 of Ref. \cite{vogel2002computational}. Tikhonov regularization fits into
the form of Eq. (\ref{reglsq}) by writing $R(\iv{f})=(1/2)\|\iv{f}\|^2_2$. One small
inconvenience with this approach, however, is that the physical units of the two
terms in the objective function of Eq. (\ref{reglsq}) are different, and therefore it can
be difficult to physically interpret the regularization parameter $\alpha$.
An equivalent optimization problem can be formulated as a special case of Eq. (\ref{cfgen})
\begin{equation}
\label{cfxfgineq}
\iv{f}^\circ = \argmin_\iv{f} \left\{ \frac{1}{2} \| \iv{f} - \iv{f}_\text{prior} \|^2_2
+ \delta_{\text{Ball}(\epsilon^\prime)}(\sm{X} \iv{f} -\iv{g})  \right\},
\end{equation}
which differs from Eq. (\ref{cfxfeg}) only in that the set $S_1$ is widened from a 0-vector
to $\text{Ball}(\epsilon^\prime)$, where
we use the term $\text{Ball}(\epsilon^\prime)$ to denote a multi-dimensional solid sphere of radius $\epsilon^\prime$ and
the dimension of the solid sphere is taken to be the same as $\text{size}(\iv{g})$.
We also define the parameter $\epsilon$, which is a constraint on the 
data RMSE
\begin{equation}
\notag
\epsilon= \epsilon^\prime/\sqrt{\text{size}(\iv{g})}.
\end{equation}
The corresponding dual maximization problem is
\begin{equation}
\label{cfxfgineqdual}
\iv{y}^\circ = \argmax_\iv{y} \left\{- \frac{1}{2} \|X^T \iv{y} \|^2_2
- \epsilon^\prime \| \iv{y} \|_2
-\iv{g}^T \iv{y} + \iv{f}^T_\text{prior} (X^T \iv{y})
\right\}.
\end{equation}
The indicator $\delta_{\text{Ball}(\epsilon^\prime)}(\sm{X} \iv{f} -\iv{g})$ 
in Eq. (\ref{cfxfgineq}) is zero when $\|\sm{X} \iv{f} -\iv{g}\|_2 \le
\epsilon^\prime$ and infinity otherwise. This optimization problem is equivalent to Tikhonov regularization
when $\iv{f}_\text{prior}$ is zero and $\epsilon^\prime >0$ in the sense that there exists a
corresponding $\alpha$ (not known ahead of time) where the two optimization problems yield the same solution.
The advantage of Eq. (\ref{cfxfgineq}) is that the parameter $\epsilon^\prime$ has a meaningful
physical interpretation as a tolerance on the data-error. Larger $\epsilon^\prime$ yields greater
regularization. Generally, the Tikhonov form is preferred due to algorithm availability.
Tikhonov regularization can be solved, for example, by linear CG. With the application
of CP2-IC, however, an accelerated solver is now available that directly solves
the constrained minimization problem in Eq. (\ref{cfxfgineq}).

\begin{figure}
\hrulefill
\begin{algorithmic}[1]
\State $L \gets \| \sm{X} \|_2; \; \tau \gets 1; \; \sigma \gets 1/L^2; \; n \gets 0$
\State initialize $\iv{f}_0$ and $\iv{y}_0$ to zero vectors
\State $\bar{\iv{f}}_0 \gets \iv{f}_0$
\Repeat
\State $\iv{y}^\prime_n \gets  \iv{y}_n+\sigma( \sm{X} \bar{\iv{f}}_n -\iv{g})$; 
$ \; \iv{y}_{n+1} \gets \max( \| \iv{y}^\prime_n\|_2 - \sigma \epsilon^\prime, 0 )
\frac{\iv{y}^\prime_n}{ \| \iv{y}^\prime_n \|_2} $ \label{dualupdate2}
\State $\iv{f}_{n+1} \gets \left[ \iv{f}_n - \tau (\sm{X}^T \iv{y}_{n+1} - \iv{f}_\text{prior})\right]
/(1+\tau)$ \label{primalupdate2}
\State $ \theta \gets 1/\sqrt{1+2\tau}$; $\tau \gets \tau \theta$; $\sigma \gets \sigma / \theta$ 
\State $\bar{\iv{f}}_{n+1} \gets \iv{f}_{n+1} + \theta(\iv{f}_{n+1} - \iv{f}_n)$
\State $n \gets n+1$
\Until{$n \ge N$}
\end{algorithmic}
\hrulefill
\caption{Pseudocode for $N$ steps of the accelerated CP algorithm instance
for solving Eq. (\ref{cfxfgineq}) with parameter $\epsilon^\prime$. Variables are defined in Sec. \ref{sec:firstalg}.}
\label{alg:cfxfgineq}
\end{figure}
The pseudocode for CP2-IC is given in
Fig. \ref{alg:cfxfgineq}.  This pseudocode differs from the previous at the update of the
dual variable $\iv{y}_{n+1}$ in Line \ref{dualupdate2}. The derivation of this dual update is covered in
detail in our previous work on the application of the CP algorithm to CT image reconstruction
\cite{SidkyCP:2012}.
For the limited angular-range CT problem considered here,
Eq. (\ref{cfxfgineq}) is particularly challenging because the constraint shape is highly
eccentric due to the spread in singular values of $\sm{X}$. 

\subsubsection{CP2-ICTV: an accelerated CP algorithm
instance for total variation and data-error constraints}

Recently, regularization based on the $\ell_1$-norm has received much attention. In particular,
the TV semi-norm has found extensive application in medical imaging due to
the fact that CT tomographic images are approximately piece-wise constant. The TV semi-norm of $\iv{f}$ is
written as $\|(| \nabla \iv{f}|) \|_1$, where $\nabla$ is a matrix encoding a finite-difference
approximation to the gradient operator; it acts on an image and yields a spatial-vector image.
The absolute value operation acts pixel-wise, taking the length of the spatial-vector at
each pixel of this image;
accordingly, $| \nabla \iv{f}|$ is the gradient-magnitude image of $\iv{f}$.
The TV semi-norm can be used as a penalty with the generic optimization problem of Eq. (\ref{reglsq}),
by setting $R(\iv{f}) = \| (|\nabla \iv{f}|) \|_1$. Convergent large-scale solvers for this optimization
problem have only recently been developed with some algorithms relying on smoothing the 
TV term \cite{Jensen2011,Defrise:11,fessler:2012}.
As with Tikhonov regularization, there is still the inconvenience
of having no physical meaning of the regularization parameter $\alpha$. We continue along the path
of recasting optimization problems as a convex feasibility problem and consider
\begin{equation}
\label{cfTVxfgineq}
\iv{f}^\circ = \argmin_\iv{f} \left\{ \frac{1}{2} \| \iv{f} - \iv{f}_\text{prior} \|^2_2
+ \delta_{\text{Ball}(\epsilon^\prime)}(\sm{X} \iv{f} -\iv{g}) + \delta_{\text{Diamond}(\gamma)}(| \nabla \iv{f}|) \right\},
\end{equation}
where the additional indicator places a constraint on the TV of $\iv{f}$; and
we have $\sm{K}_1(\iv{f}) = \sm{X} \iv{f} +\iv{g}$, $\sm{K}_2(\iv{f}) = \nabla \iv{f}$,
$S_1= \{\iv{g} \text{ such that } \iv{g} \in \text{Ball}(\epsilon^\prime)$ \}, and
$S_2 = \{\iv{z} \text{ such that } |\iv{z}| \in \text{Diamond}(\gamma) \}$, where $\iv{z}$
is a spatial-vector image.
The term $\text{Diamond}(\gamma)$
describes the $\ell_1$-ball of scale $\gamma$; the indicator $\delta_{\text{Diamond}(\gamma)}(| \nabla \iv{f}|)$
is zero when $\|(| \nabla \iv{f}|) \|_1 \le \gamma$. This convex feasibility problem asks for the image
that is closest to $\iv{f}_\text{prior}$ and satisfies the $\epsilon^\prime$-data-error and $\gamma$-TV-constraints.
The corresponding dual maximization problem is
\begin{equation}
\label{cfTVxfgineqdual}
\iv{y}^\circ = \argmax_{\iv{y}, \iv{z}}
\left\{- \frac{1}{2} \|X^T \iv{y} + \nabla^T \iv{z} \|^2_2
- \epsilon^\prime \| \iv{y} \|_2 - \gamma \| (|\iv{z}|) \|_{\infty}
-\iv{g}^T \iv{y} + \iv{f}^T_\text{prior} (X^T \iv{y}+ \nabla^T \iv{z})
\right\},
\end{equation}
where $\iv{z}$ is a spatial-vector image; $|\iv{z}|$ is the scalar image
produced by taking the vector magnitude of $\iv{z}$ at each pixel;
the $\ell_{\infty}$-norm yields the largest component of the vector argument;
and $\nabla^T$ is the matrix transpose of $\nabla$.
We demonstrate in Sec. \ref{sec:results} application of CP2-ICTV to both inconsistent and consistent constraint
sets.
Due to the length of the pseudocode, we present it in Appendix \ref{sec:CF3-CP}, and
point out that it can be derived following
Ref. \cite{SidkyCP:2012}, using the Moreau identity described in Ref. \cite{chambolle2011first}
and an algorithm for projection onto the $\ell_1$-ball \cite{duchi2008efficient}.

\subsection{Summary of proposed convex feasibility methodology}

Our previous work in Ref. \cite{SidkyCP:2012} promoted use of CP Algorithm 1
to prototype convex optimization problems for IIR in CT. Here, we
restrict the convex optimization problems to the form of Eq. (\ref{cfgen}),
allowing the use of the accelerated CP Algorithm 2 with a steeper worst-case convergence rate.
Because the proposed optimization problem Eq. (\ref{cfgen}) has a generic convex feasibility term,
the framework can be regarded as convex feasibility prototyping.
The advantage of this approach is two-fold:
(1) an accelerated CP algorithm is available with an $O(1/N^2)$ convergence
rate, and
(2) the design of convex feasibility connects better with physical
metrics related to the image estimate.
To appreciate the latter point, consider the unconstrained counterpart
to ICTV. In setting up an objective function which is the sum of image TV, data fidelity,
and distance from $\iv{f}_\text{prior}$, two parameters are needed to balance
the strength of the three terms. We arrive at
\begin{equation}
\notag
\iv{f}^\circ = \argmin_\iv{f} \left\{
\frac{1}{2} \| \iv{f} - \iv{f}_\text{prior} \|^2_2 +
 \alpha_1 \frac{1}{2} \| \iv{g} - \sm{X} \iv{f} \|^2_2 +
\alpha_2 \|\iv{f}\|_{TV} \right\}.
\end{equation}
As the terms reflect different physical properties
of the image, it is not clear at all what values should be selected nor
is it clear what the impact of the parameters are on the solution of the unconstrained
minimization problem.

The following results section demonstrates use of CP2-EC, CP2-IC,
and CP2-ICTV on a breast CT simulation with a limited scanning angular range.
The main goals of the numerical examples are to demonstrate use of the 
proposed convex feasibility framework and convergence properties of the
derived algorithms. Even though the algorithms are known to converge within
a known worst-case convergence rate, it is still important to observe
the convergence of particular image metrics
in simulations similar to an actual application.

\section{Results: demonstration of the convex feasibility accelerated CP algorithms}
\label{sec:results}

\begin{figure}[!h]
\begin{minipage}[b]{\linewidth}
\centering
\centerline{\includegraphics[width=0.6\linewidth]{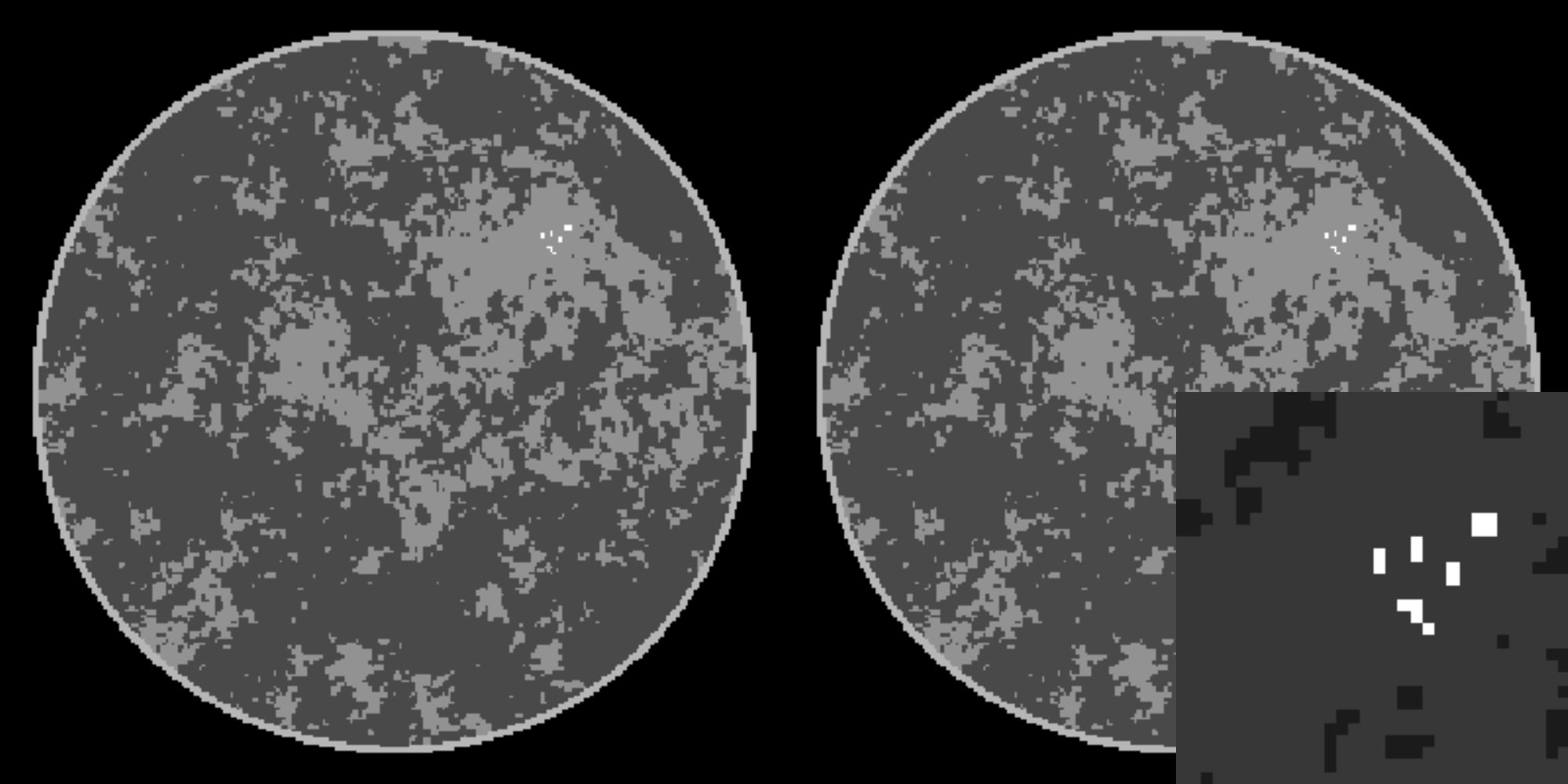}}
\end{minipage}
\caption{Breast phantom for the CT limited angular-range scanning
simulation.
Left: the phantom in the gray scale window [0.95,1.15]. Right: the same
phantom with a blow-up on the micro-calcification ROI displayed in the gray
scale window [0.9,1.8].
The right panel is the reference for all image reconstruction algorithm results.
\label{fig:phantom}}
\end{figure}
We demonstrate the application of the various accelerated CP algorithm instances on simulated
CT data generated from the breast phantom shown in Fig. \ref{fig:phantom}.
The phantom, described in
Ref. \cite{joergensen2011toward,reiser2010task}, is digitized on a 256 $\times$ 256 pixel array.
Four tissue types are modeled:
the background fat tissue is taken as the reference material
and assigned a value of 1.0,
the modeled fibro-glandular tissue takes a value of 1.1, the outer skin layer is set to 1.15,
and the micro-calcifications are assigned values in the range [1.8,2.3]. The simulated CT
configuration is described at the beginning of Sec. \ref{sec:methods}.

In the following, the IIR algorithms are demonstrated with ideal data generated by applying
the system matrix $\sm{X}$ to the phantom and with inconsistent data obtained by adding Poisson
distributed noise to the ideal data set. We emphasize that the goal of the paper is to address
convergence of difficult optimization problems related to IIR in limited angular-range CT.
Thus, we are more interested in establishing that the CP algorithm instances achieve accurate
solution to their corresponding optimization problems, and we are less concerned about 
the image quality of the reconstructed images. In checking convergence in the consistent
case, we monitor the conditional primal-dual gap.

For the inconsistent case, we do not have a general criterion for convergence.
The conditional primal-dual gap tends to infinity because the dual objective function
is forced to tend
to infinity in order to meet the primal objective function, which is necessarily infinity
for inconsistent constraints.
We
hypothesize, however, that
CP2-EC minimizes the least-squares problem, Eq. (\ref{lsq}), and we can use the gradient magnitude of
the least-squares objective function
to check this hypothesis and test convergence. For CP2-IC, we also hypothesize
that it solves
the same problem in the inconsistent case, but it is not interesting because we can instead use
the parameter-less EC problem. Finally, for CP2-ICTV
we do not have a convergence check in the
inconsistent case, but we also note that it is difficult to say whether or not a specific
instance of ICTV is consistent or not because there are two constraints on quite different image metrics.
For this problem the conditional primal-dual gap is useful for making this determination.
If we observe a divergent trend in the conditional primal-dual gap, we can say that the particular
choice of TV and data-error constraints are not compatible.

Additionally, we monitor two other metrics as a function of iteration number,
the image RMSE is
\begin{equation}
\notag
\frac{ \| \iv{f} - \iv{f}_\text{phantom} \| _2} {\sqrt{ \text{size}(\iv{f}) }},
\end{equation}
and the data RMSE is
\begin{equation}
\notag
\frac{ \| \iv{g} - \sm{X} \iv{f} \| _2} {\sqrt{ \text{size}(\iv{g}) }}.
\end{equation}
We take the former as a surrogate for image quality, keeping in mind the pitfalls
in using this metric, see Sec. 14.1.2 of Ref. \cite{Barrett:FIS}. The latter
along with image TV are used to verify that the constraints are being satisfied.

\subsection{Ideal data and equality-constrained optimization}

We generate ideal data from the breast phantom and apply CP2-EC, with $\iv{f}_{prior}=0$, to investigate
its convergence behavior for limited angular-range CT. As the simulations is set up 
so that $\sm{X}$ is left-invertible and the data are generated from applying this system
matrix to the test phantom, the indicator $\delta_0(\sm{X} \iv{f}-\iv{g})$
in Eq. (\ref{cfxfeg}) is zero only when $\iv{f}$ is the phantom. Observing convergence
to the breast phantom as well as the rate of convergence is of main interest here.

In order to have a reference to standard algorithms, we apply linear CG \cite{Nocedal:06} and
ART to the same problem.  Linear CG solves
the minimization problem in Eq. (\ref{lsq}), which corresponds to solving
the linear system in Eq. (\ref{xfegConsistent}).
The matrix, $\sm{X}^T \sm{X}$, in this equation is symmetric with non-negative singular values.
The ART algorithm, which is a form of POCS, solves Eq. (\ref{xfeg}) directly by cycling through orthogonal
projections onto the hyper-planes specified by each row of the linear system.

\begin{figure}[!h]
\begin{minipage}[b]{0.45\linewidth}
\centering
\centerline{\includegraphics[width=\linewidth]{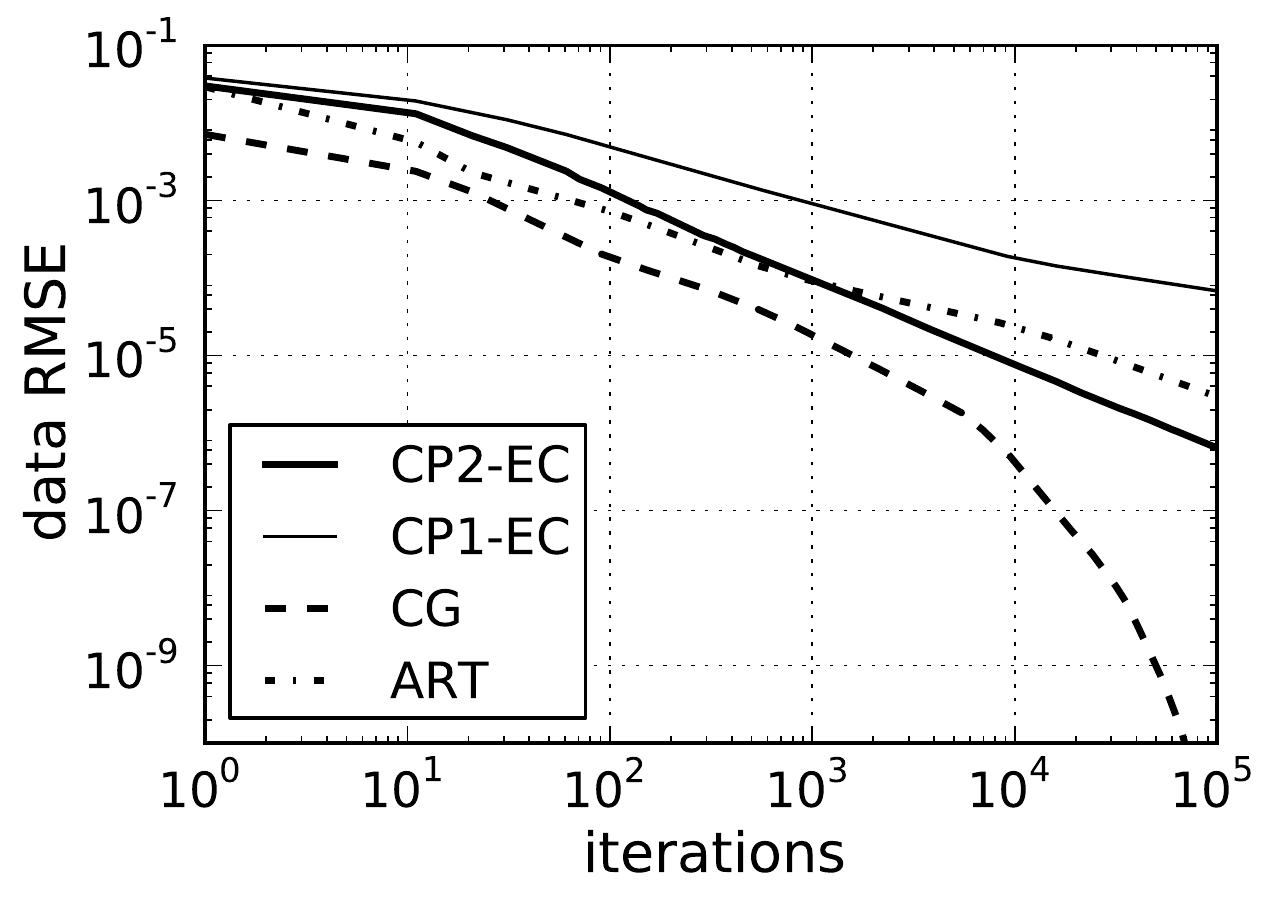}}
\end{minipage}
\begin{minipage}[b]{0.45\linewidth}
\centering
\centerline{\includegraphics[width=\linewidth]{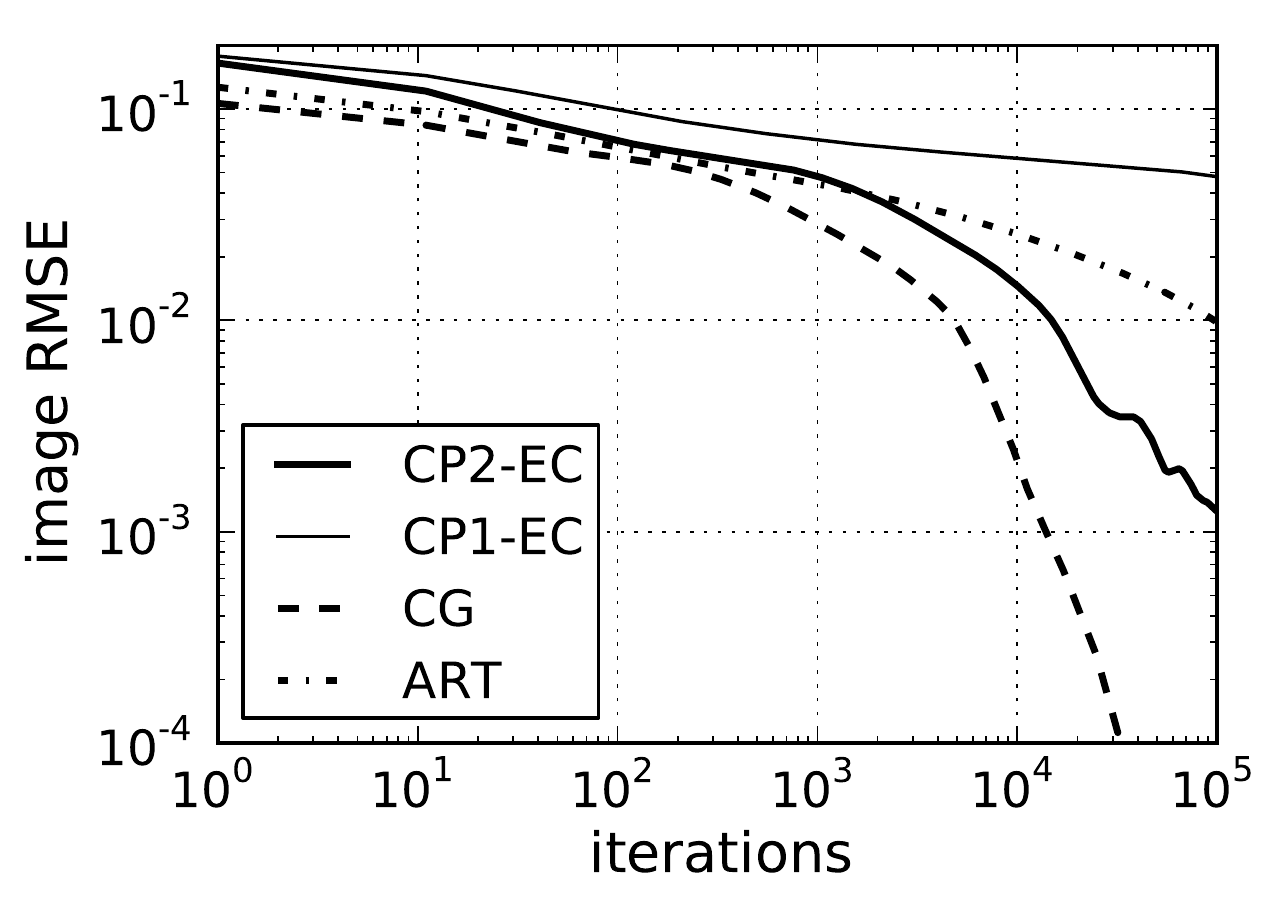}}
\end{minipage}
\begin{minipage}[b]{\linewidth}
\centering
\centerline{CG~~~~~~~~~~~~~~~~~~~~~~~~~~~~~~~~~~~ART}
\vskip 0.1cm
\centerline{\includegraphics[width=0.6\linewidth]{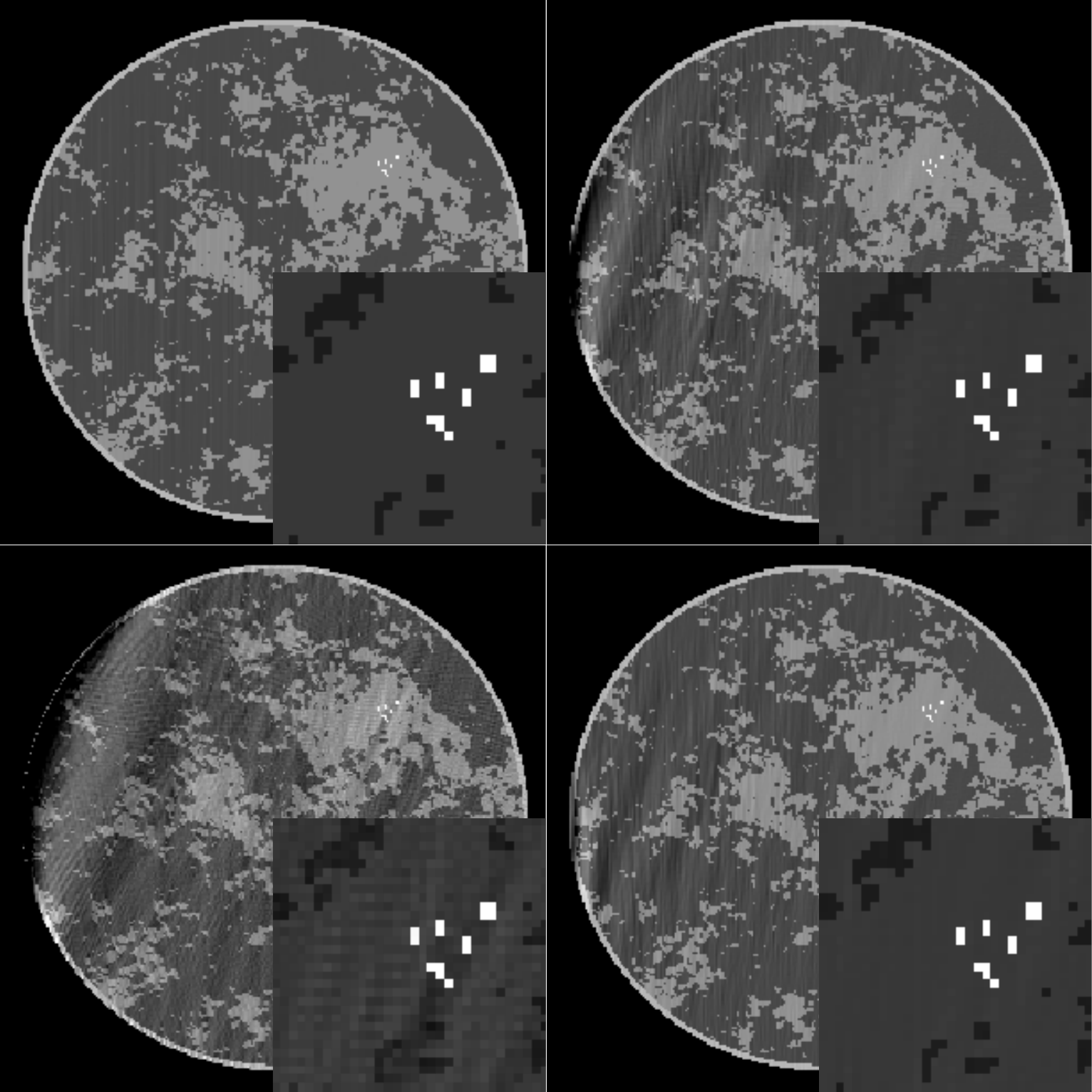}}
\vskip -0.3cm
\centerline{CP1-EC~~~~~~~~~~~~~~~~~~~~~~~~~~~~~~~CP2-EC}
\end{minipage}
\caption{Results of CP2-EC 
with ideal, simulated data. Convergence is also compared
with CP1-EC, linear CG,
and ART. Top row: (Left) convergence of the four algorithms in terms of data RMSE, and
(Right) convergence of the four algorithms in terms of image RMSE.
Bottom row: the image at iteration 10000 for CG, ART, CP Algorithm 1, and CP2-EC
shown in the same gray scale as Fig. \ref{fig:phantom}.
The artifacts seen at the right of the images and relatively
large image RMSE are indications of the poor
conditioning of $\sm{X}$. The comparison between CP2-EC and CP1-EC
shows quantitatively the impact of the acceleration afforded by CP Algorithm 2.
\label{fig:idealData}}
\end{figure}

The results of each algorithm are shown in Fig. \ref{fig:idealData}. As the data are ideal, each algorithm
drives the data-error to zero.  The linear CG algorithm shows the smallest data RMSE, but
we note similar slopes on the log-log plot of CG and CP2-EC during most of the computed
iterations except near the end, where the slope of the CG curve steepens.
The ART algorithm reveals a convergence slightly
faster than CP2-EC, initially, but it is overtaken by CP2-EC near iteration 1000.
We also note the impact of the algorithm acceleration afforded by the proposed convex feasibility
framework in the comparison of CP2-EC and CP1-EC.

Because $\sm{X}$ is designed to be left-invertible, we know also that the image estimates must
converge to the breast phantom
for each of the four algorithms.  A similar ordering of the convergence rates is observed in the
image RMSE plot, but we note that the values of the image RMSE are all much larger than corresponding
values in the data RMSE plots. This stems from the poor conditioning of $\sm{X}$, and this point is
emphasized in examining the shown image estimates at iteration 10000 for each algorithm.

While the image RMSE gives a summary metric on the accuracy of the image reconstruction,
the displayed images yield more detailed information
on the image error incurred by truncating the algorithm iteration. The CP2-EC, CP1-EC, and ART
images show
wavy artifacts on the left side; the limited-angle scanning arc is over the right-side
of the object. But the CG image shows visually accurate image reconstruction
at the given gray scale window setting.

This initial result shows promising convergence rates for CP2-EC and that it may
be competitive with existing algorithms for solving large, consistent linear systems. But we cannot draw any
general conclusions on algorithm convergence,  because different simulation conditions
may yield different ordering of the convergence rates.  Moreover, we have implemented
only the basic forms of CG and ART; no attempt at pre-conditioning CG was made and
the relaxation parameter of ART was fixed at 1.

\begin{figure}[!h]
\begin{minipage}[b]{0.45\linewidth}
\centering
\centerline{\includegraphics[width=\linewidth]{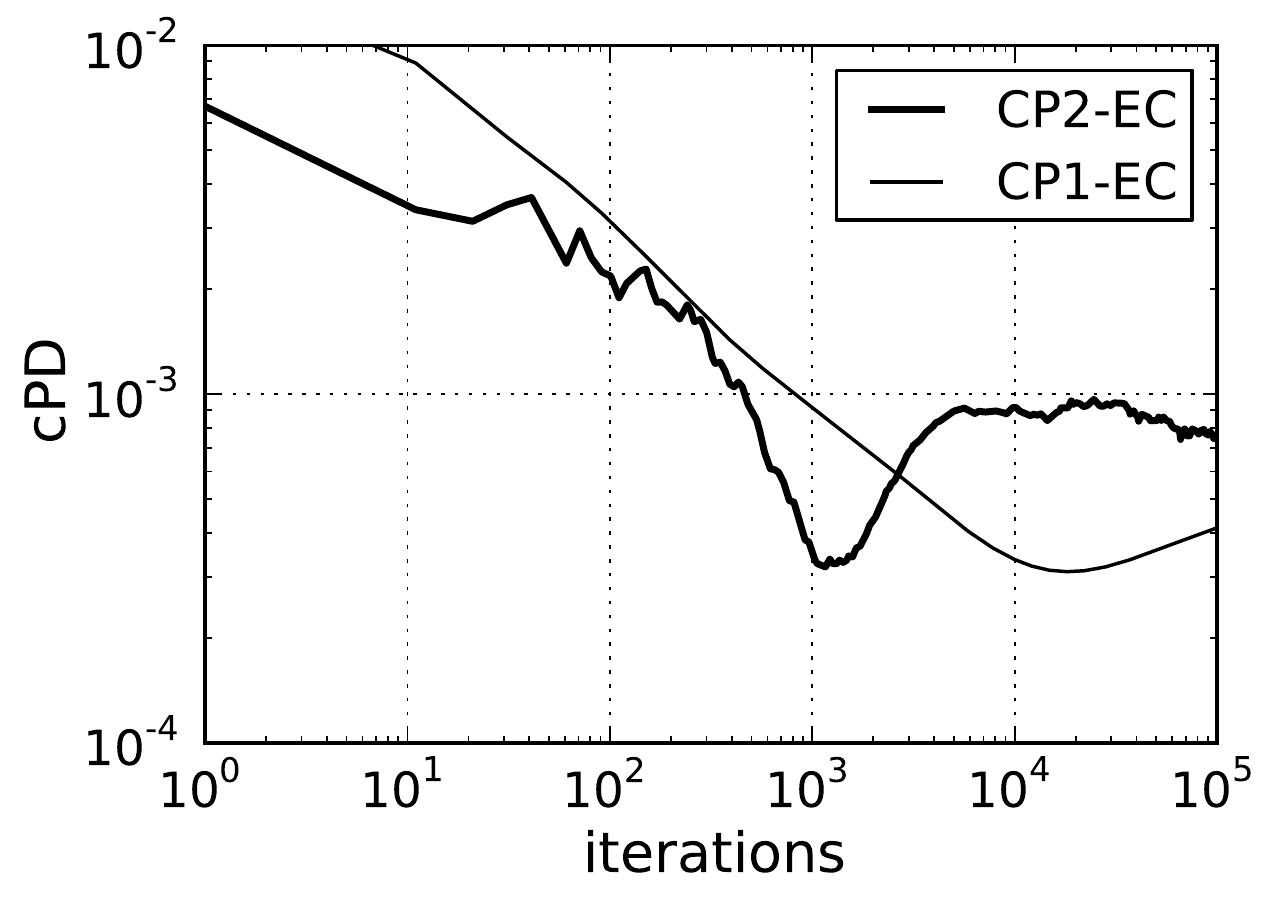}}
\end{minipage}
\caption{The conditional primal-dual gap for EC shown for CP2-EC and
CP1-EC. This gap is computed
by taking the difference between the primal and dual objective functions in Eqs. (\ref{cfxfeg})
and (\ref{cfxfegdual}), respectively, after removing the indicator in the
primal objective function:
$cPD= \left|\frac{1}{2} \| \iv{f} - \iv{f}_\text{prior} \|^2_2 +
\frac{1}{2} \|X^T \iv{y} \|^2_2
+\iv{g}^T \iv{y} - \iv{f}^T_\text{prior} (X^T \iv{y})
\right|/\text{size}(\iv{f})$. The absolute value is used because the argument can be negative,
and we normalize by the number of pixels $\text{size}(\iv{f})$ so that the primal objective function
takes the form of a mean square error. The prior image $\iv{f}_\text{prior}$  for this 
computation is zero. The comparison between CP2-EC and CP1-EC
shows quantitatively the impact of the acceleration afforded by CP Algorithm 2.
\label{fig:idealDataGap}}
\end{figure}

We discuss convergence in detail as it is a major focus of this article.
In Fig. \ref{fig:idealDataGap}, we display the conditional primal-dual gap for the
accelerated CP2-EC algorithm compared with use of CP1-EC. First, it is clear that
convergence of this gap is slow for this problem due to the ill-conditionedness of $\sm{X}$,
and we note this slow convergence is in line
with the image RMSE curves in Fig. \ref{fig:idealData}. The image RMSE has
reached only $10^{-3}$ after 10$^5$ iterations. 
Second, the gap for CP1-EC appears to be lower than that of CP2-EC at the final
iteration, but the curve corresponding to CP2-EC went through a similar dip and is returning
to a slow downward trend. Third, for a complete convergence check, we must examine the constraints
separately from the conditional primal-dual gap, The only constraint in EC is formulated
in the indicator $\delta_0(\sm{X} \iv{f}-\iv{g})$. In words, this constraint is that the given
data and data estimate must be equal or, equivalently, the data RMSE must be zero. We observe
in Fig. \ref{fig:idealData} that the data RMSE is indeed tending to zero. Now that we
have a specific example, we reiterate the
need for dividing up the convergence check into the conditional primal-dual gap and separate
constraint checks. Even though the data RMSE is tending to zero, it is not numerically
zero at any iteration and consequently the value of $\delta_0(\sm{X} \iv{f}-\iv{g})$ is
$\infty$ at all iterations. Because this indicator is part of the primal objective function in Eq. (\ref{cfxfeg}),
this objective function also takes on the value of $\infty$ at all iterations. As a result, direct
computation of the primal-dual gap does not provide a useful convergence check and we need
to use the conditional primal-dual gap.


\subsection{Noisy, inconsistent data and equality-constrained optimization}

In this section, we repeat the previous simulation with all four algorithms
except that the data now contain inconsistency modeling Poisson distributed noise.
The level of the noise is selected to simulate what could be seen in a low-dose CT scan.
The use of this data model contradicts the application of equality-constrained optimization
and EC becomes inconsistent.
But nothing prevents us
from executing the CP2-EC operations, and accordingly we do so in this
subsection. The linear CG algorithm can still be applied in this case, because
the optimization problem in Eq. (\ref{lsq}) is well-defined even though there
is no $\iv{f}$ such that $\iv{g} = \sm{X} \iv{f}$. Likewise, the linear system
in Eq. (\ref{xfegConsistent}) does have a solution even when $\iv{g}$ is inconsistent.
The basic ART algorithm, as with CP2-EC, is not suited to this data model, because it
is a solver for Eq. (\ref{xfeg}), which we know ahead of time has no solution.
Again, as with CP2-EC, the steps of ART can still be executed even with inconsistent data,
and we show the results here.

\begin{figure}[!h]
\begin{minipage}[b]{0.45\linewidth}
\centering
\centerline{\includegraphics[width=\linewidth]{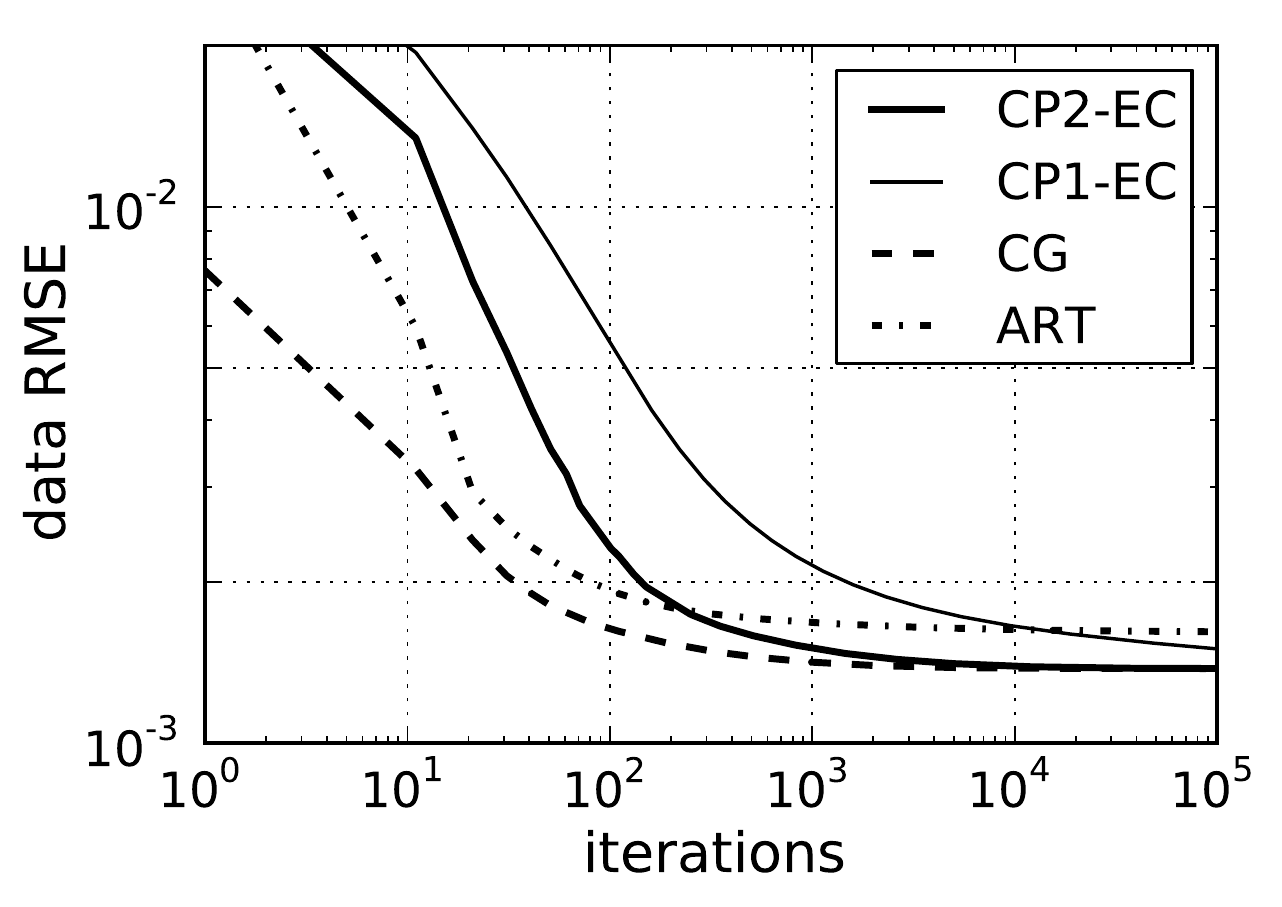}}
\end{minipage}
\begin{minipage}[b]{0.45\linewidth}
\centering
\centerline{\includegraphics[width=\linewidth]{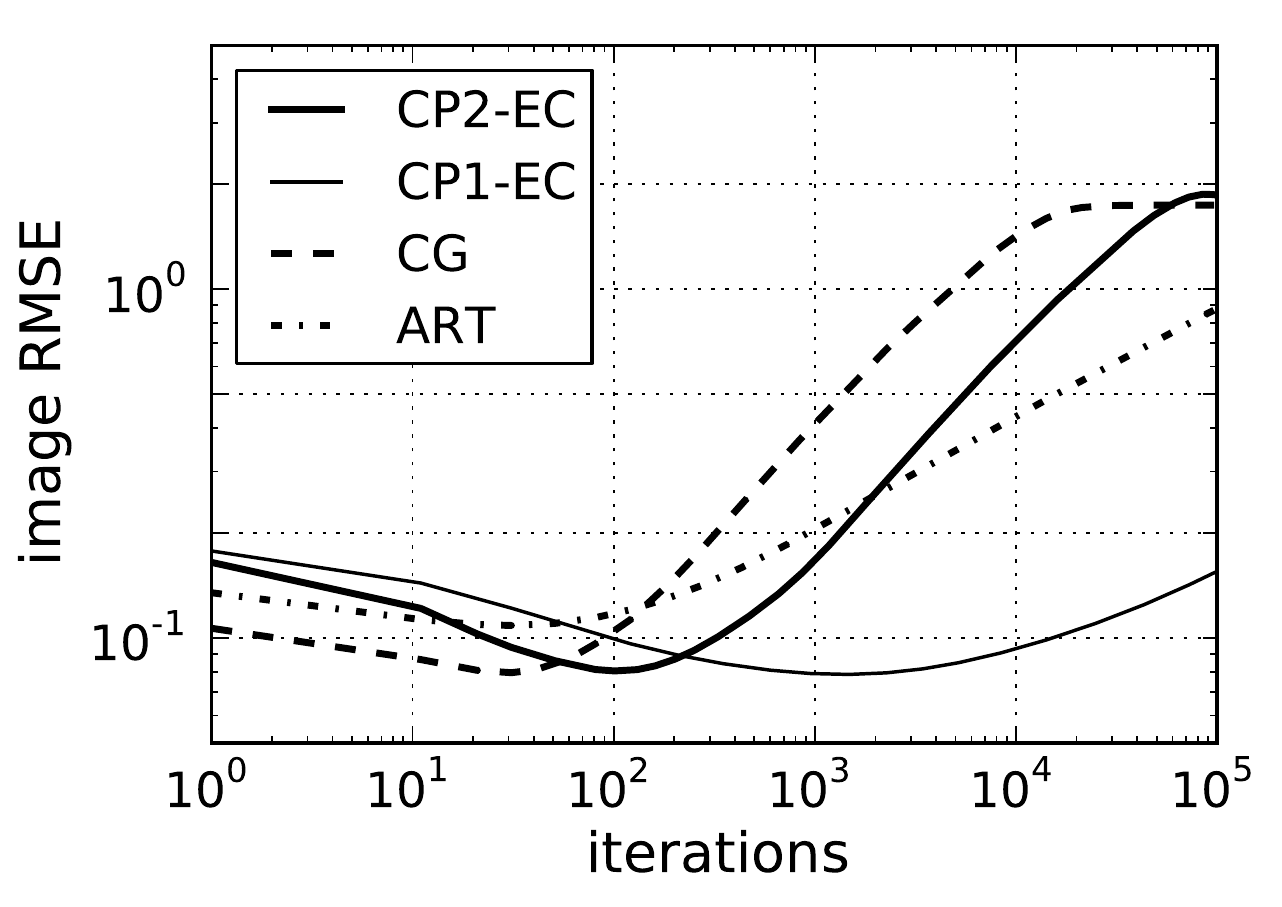}}
\end{minipage}
\caption{Metrics of CP2-EC image estimates
with noisy and inconsistent, simulated data. Results are compared
with CP1-EC, linear CG, and ART.
Left, evolution of the four algorithms in terms of data RMSE,
and
right, evolution of the four algorithms in terms of image RMSE.
\label{fig:noisyDataEq}}
\end{figure}

In Fig. \ref{fig:noisyDataEq}, we show evolution plots of quantities
derived from the image estimates from each of the four algorithms.
Because the data are inconsistent, the data- and image-error plots have a different behavior
than the previous consistent example.
In this case, we know that the data RMSE cannot
be driven to zero. The algorithms CP2-EC and CG converge on a value
greater than zero,
while CP1-EC and ART appear to need more iterations to reach the same data RMSE value.

The image RMSE shows an initial decrease to some minimum value followed by
an upward trend. For CG the upward trend begins to level off at 20,000 iterations,
while for CP2-EC it appears that this happens near the
final 100,000th iteration. For both plots, the results of CP1-EC lag those of the
accelerated CP2-EC algorithm.

\begin{figure}[!h]
\begin{minipage}[b]{0.45\linewidth}
\centering
\centerline{\includegraphics[width=\linewidth]{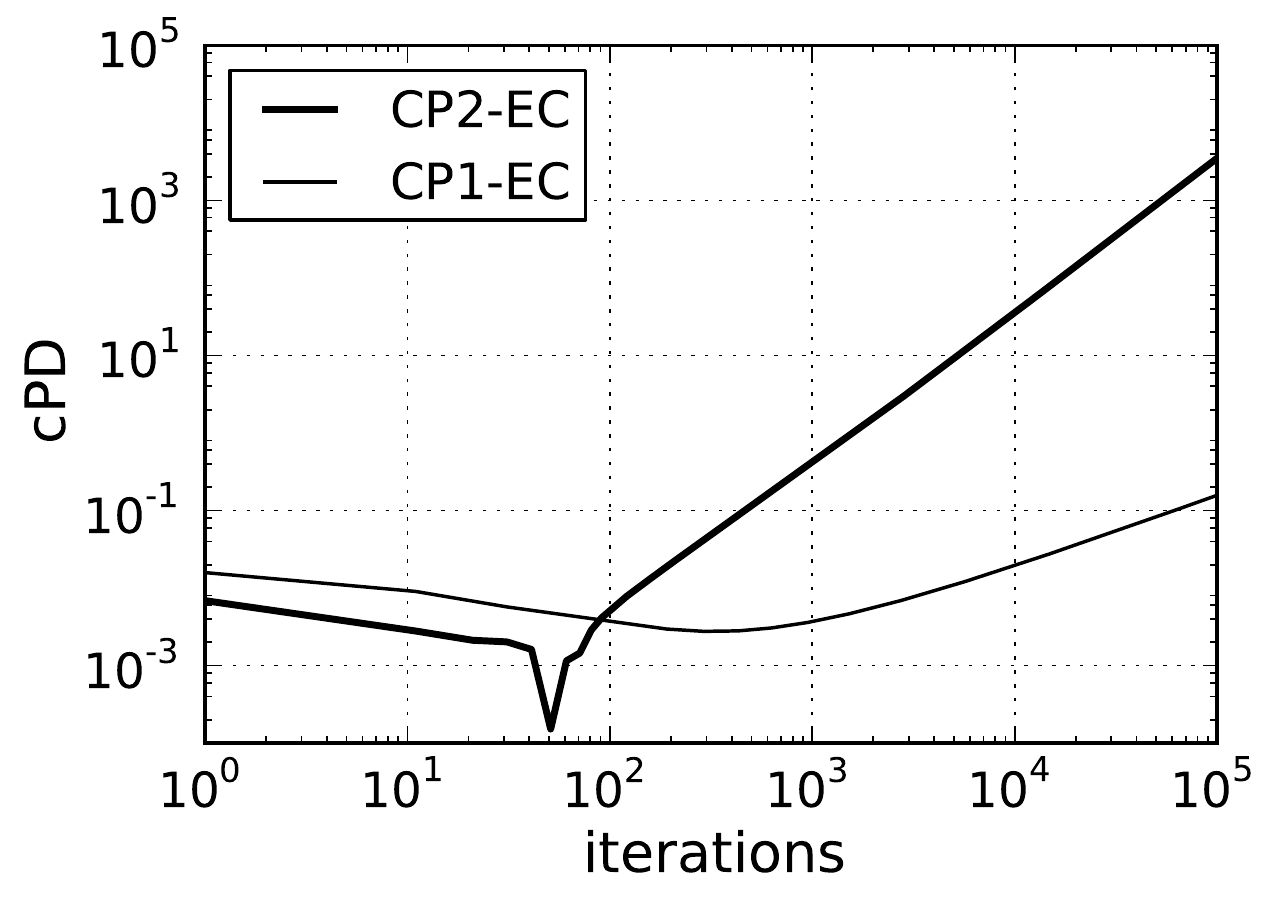}}
\end{minipage}
\begin{minipage}[b]{0.45\linewidth}
\centering
\centerline{\includegraphics[width=\linewidth]{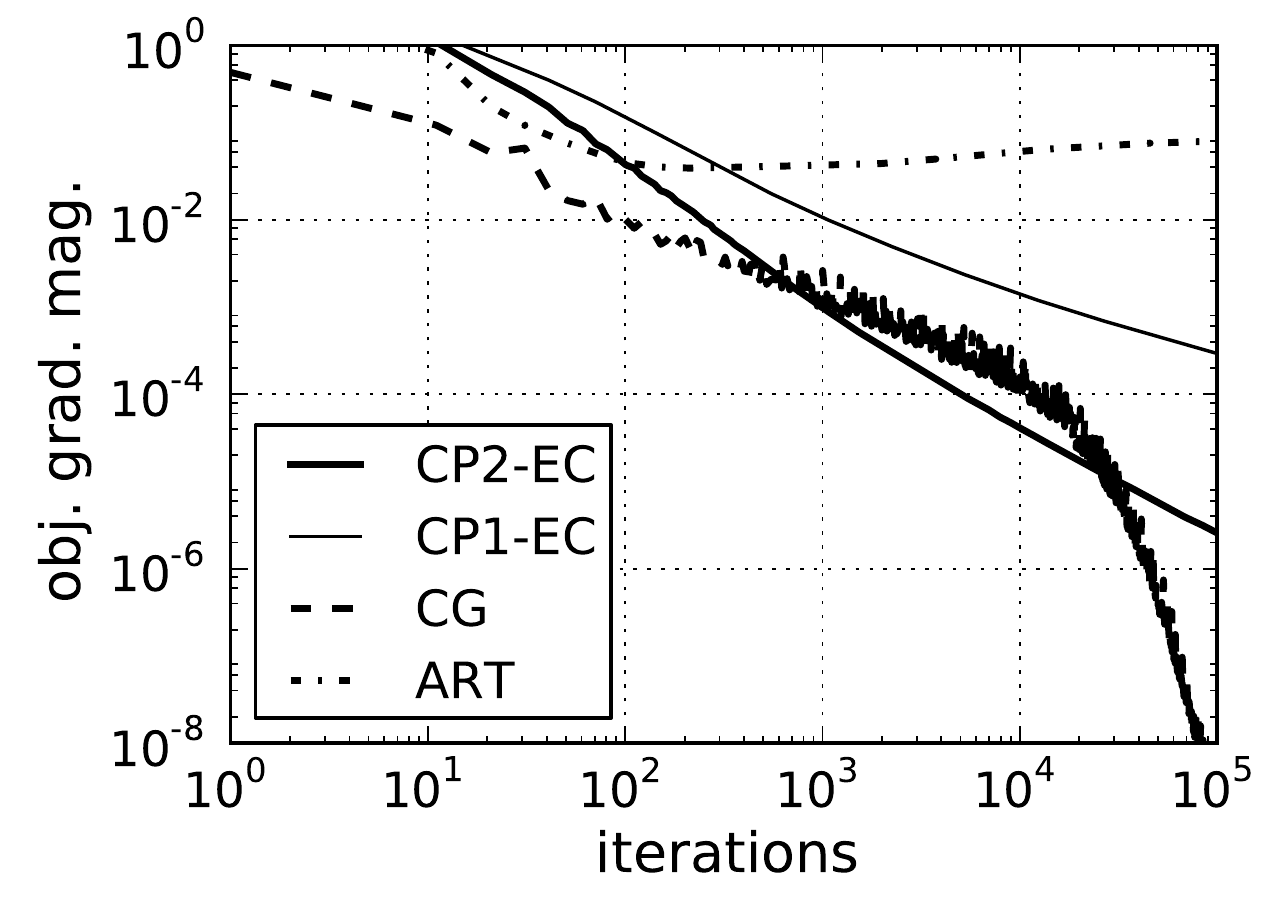}}
\end{minipage}
\caption{Convergence plots: the conditional primal-dual gap for EC (left)
and the gradient magnitude of the quadratic least-squares objective function of Eq. (\ref{lsq})
(right). The conditional primal-dual gap is only available for CP2-EC and
CP1-EC, while all algorithms can be compared with the objective function
gradient. The quantity cPD for this problem is explained in the caption
of Fig. \ref{fig:idealDataGap}. The convex feasibility problem EC is inconsistent
for the simulated noisy data, and as a result cPD diverges to $\infty$.
We hypothesize that CP2-EC converges the least squares minimization problem Eq. (\ref{lsq}),
and indeed we note in the gradient plot that CP2-EC yields a decaying
objective function gradient-magnitude competitive with linear CG and ART.
The comparison between CP2-EC and CP1-EC
shows quantitatively the impact of the acceleration afforded by CP Algorithm 2.
\label{fig:noisyDataGap}}
\end{figure}

Turning to convergence checks, we plot the conditional primal-dual gap for EC and
the magnitude of the gradient of the least-squares
objective function from Eq. (\ref{lsq}) in Fig. \ref{fig:noisyDataGap}.
As explained at the beginning of Sec. \ref{sec:results}, the conditional primal-dual
gap tends to infinity for inconsistent convex feasibility problems because the dual
objective function increases without bound.
We observe, in fact, that
the conditional primal-dual gap for EC is diverging - a consequence
of the inconsistent data used in this simulation.
In examining the objective function gradient magnitude,
the curve for the CG results shows an overall convergence by
this metric, because this algorithm is designed to solve
the normal equations of the unregularized, least-squares problem in Eq. (\ref{lsq}).
The ART algorithm shows an initial decay followed by a slow increase. This result
is not surprising, because ART is designed to solve Eq. (\ref{xfeg}) directly and not
the least-squares minimization problem in Eq. (\ref{lsq}).
As an aside, we point out that in applying ART to inconsistent data it is important
to allow the relaxation parameter to decay to zero.
Interestingly, CP2-EC and CP1-EC show a monotonic decrease of this gradient.

The resulting gradient magnitude curves indicate convergence
of the least-squares minimization, obtained by the
CP algorithms. This is surprising, because the conditional
primal-dual gap diverges to infinity.
Indeed, the magnitude of the dual variable $\iv{y}_n$
from the algorithm listed in Fig. \ref{alg:cfxfeg} increases
steadily with iteration number.  Even though the dual problem diverges, this simulation
indicates convergence of the primal
least-squares minimization problem in that the gradient of this
objective function is observed to monotonically decrease.
There is no proof that we are aware of, which covers
this situation, thus we cannot claim that CP2-EC will always converge the least-squares problem.
Therefore, in applying CP2-EC in this way it is crucial to evaluate the convergence
criterion and to verify that the magnitude of the objective function's gradient decays to zero.
The conditional primal-dual gap cannot be used as a check for CP2-EC applied to inconsistent data.

The dependence of the gradient magnitude
of the unregularized, least-squares objective function for the CP2-EC
and CG algorithms is quite interesting. Between 10 and 20,000 iterations, CP2-EC
shows a steeper decline in this convergence metric. But greater than 20,000 iterations
the CG algorithm takes over and this metric drops precipitously.  The CG behavior
can be understood in realizing that the image has approximately 50,000 unknown
pixel values and if there is no numerical error in the calculations, the CG
algorithm terminates when the number of iterations equals the number of unknowns.
Because numerical error is present, we do not observe exact convergence when
the iteration number reaches 50,000, but instead the steep decline in
the gradient of the least-squares objective function is observed. This comparison
between CP2-EC and CG has potential implications for larger systems
where the steep drop-off for CG would occur at higher iteration number.

The conditions of this particular simulation are not relevant to practical application
because it is already well-known that minimizing unregularized, data-fidelity objective functions
with noisy data converges to an extremely noisy image particularly for
an ill-conditioned system matrix; noting the large values of
the image RMSE, we know this to be the case without displaying the image. But this
example is interesting in investigating convergence properties. While it is true
that monitoring the gradient magnitude of the least-squares objective function yields
a sense about convergence, we do not know {\it a priori} what threshold
this metric needs to cross before we can say the IIR is converged, see Ref.
\cite{Nonuniform_MIC} for further discussion on this point related to IIR in CT.
This example in particular highlights the point that an image metric
of interest, such as image and data RMSE, needs to be observed to level off
in combination with a steady decrease of a convergence metric.
For this example, convergence of the image RMSE occurs when the gradient-magnitude of the
least-squares objective function drops below $10^{-5}$, while the data RMSE convergence
occurs earlier.

\subsection{Noisy, inconsistent data with inequality-constrained optimization}

In performing IIR with inconsistent projection data, some form of regularization
is generally needed. In using the convex feasibility approach, we apply CP2-IC
after deciding on the parameter $\epsilon^\prime$. The parameter $\epsilon^\prime$ has a minimum
value, below which no images satisfy the data-error constraint, and larger $\epsilon^\prime$
leads to greater image regularity.
The choice of $\epsilon^\prime$ may be guided by
properties of the available data or a prior reconstruction.
In this case, we have results from the previous section and we note that the data RMSE
achieve values below 0.002. Accordingly, for the present simulation
we select a tight data-error constraint $\epsilon^\prime = 0.512$, which is
equivalent to allowing
a data RMSE of $\epsilon=0.002$. The CP2-IC algorithm selects the image obeying
the data-error constraint closest to $\iv{f}_\text{prior}$, and to illustrate the dependence
on $\iv{f}_\text{prior}$ we present results for two choices: an image of zero values,
and an image set to 1 over the support of the phantom. Note that the second choice 
assumes prior knowledge of the object support and background value of 1.
To our knowledge, there is no direct, existing algorithm for solving
Eq. (\ref{cfxfgineq}), and thus we display results for CP2-IC only.
One can use a standard
algorithm such as linear CG to solve the Lagrangian form of Eq. (\ref{cfxfgineq}), but this
method is indirect because it is not known ahead of time what Lagrange multiplier leads
to the desired value of $\epsilon^\prime$.

\begin{figure}[!h]
\begin{minipage}[b]{0.45\linewidth}
\centering
\centerline{\includegraphics[width=\linewidth]{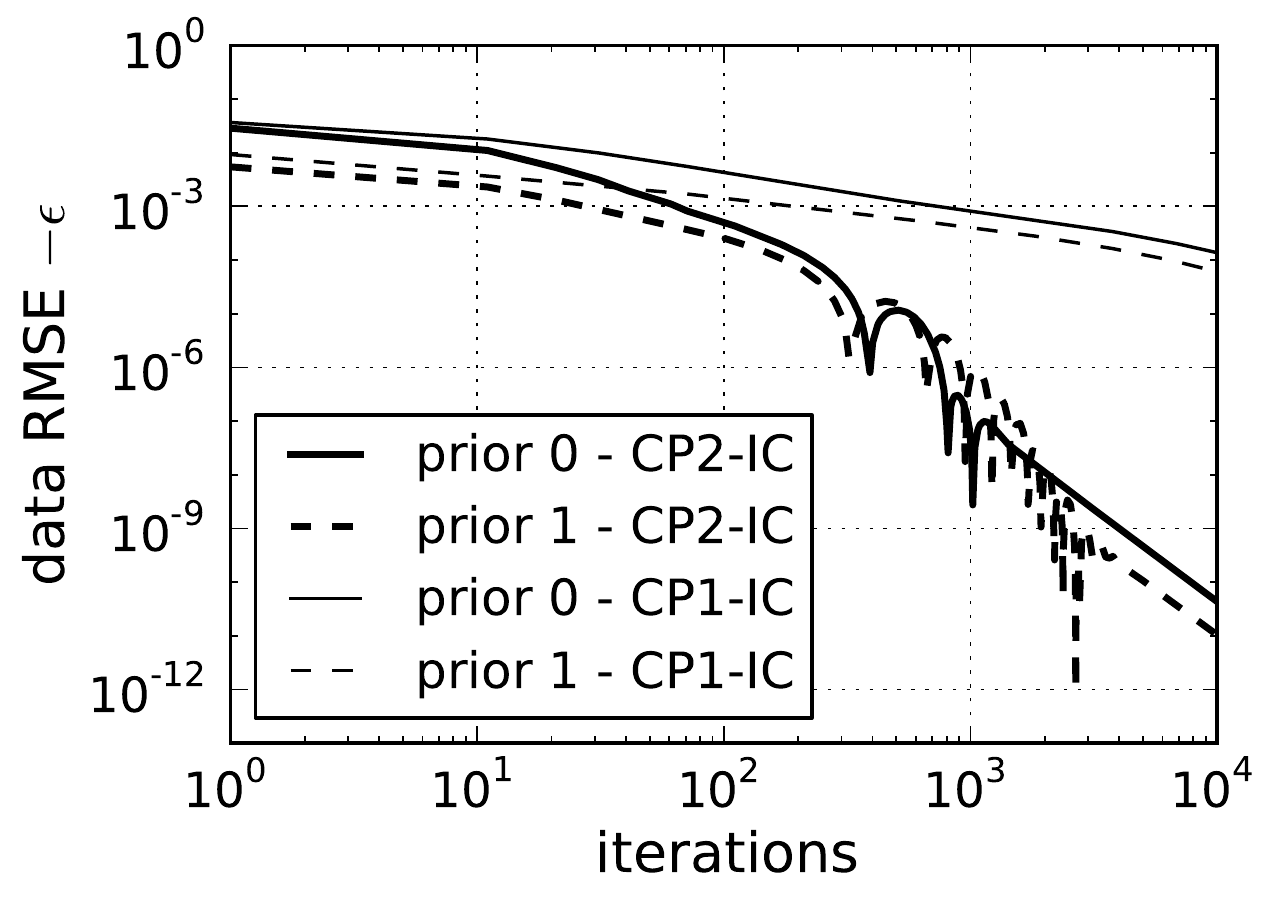}}
\end{minipage}
\begin{minipage}[b]{0.45\linewidth}
\centering
\centerline{\includegraphics[width=\linewidth]{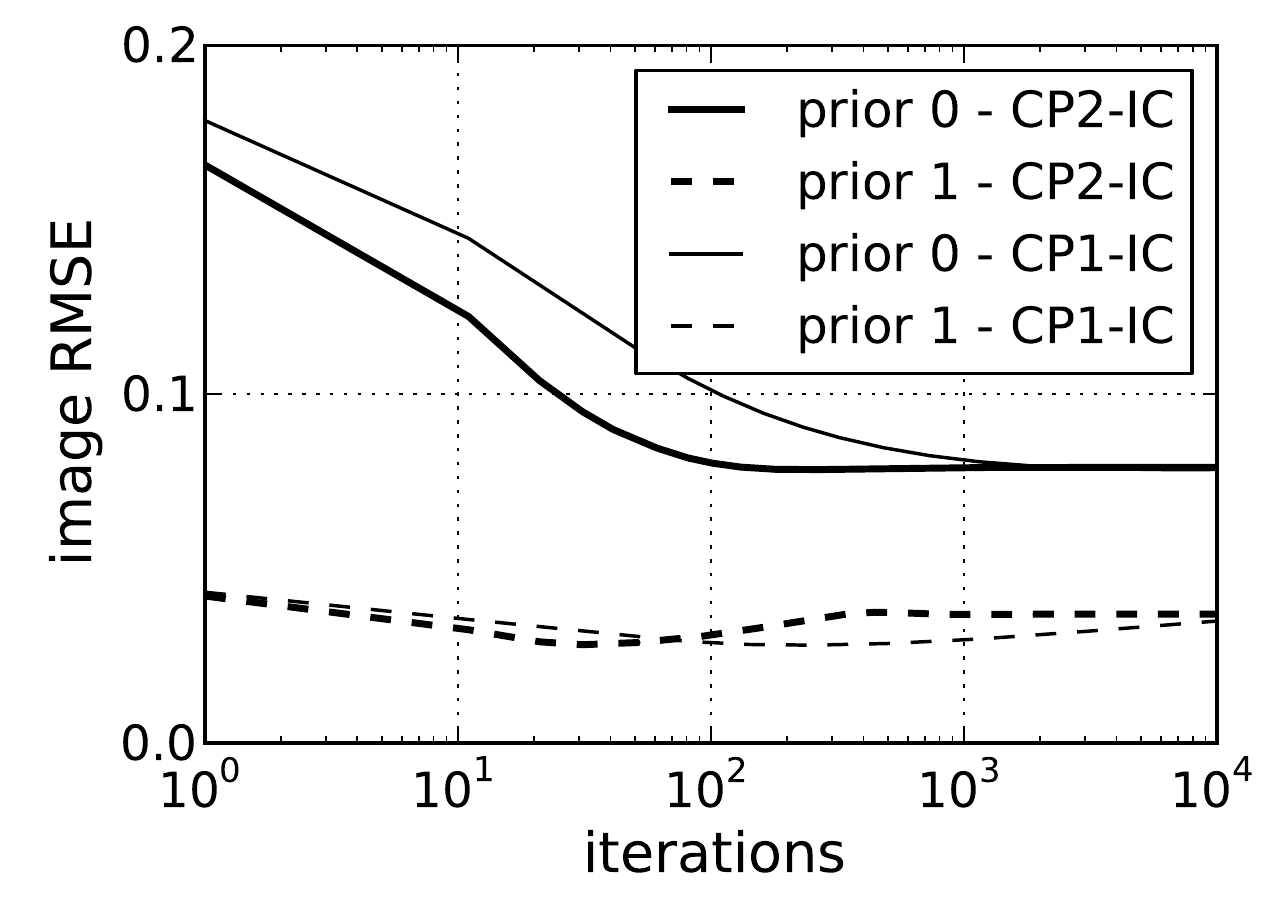}}
\end{minipage}
\begin{minipage}[b]{\linewidth}
\centering
\centerline{\includegraphics[width=0.7\linewidth]{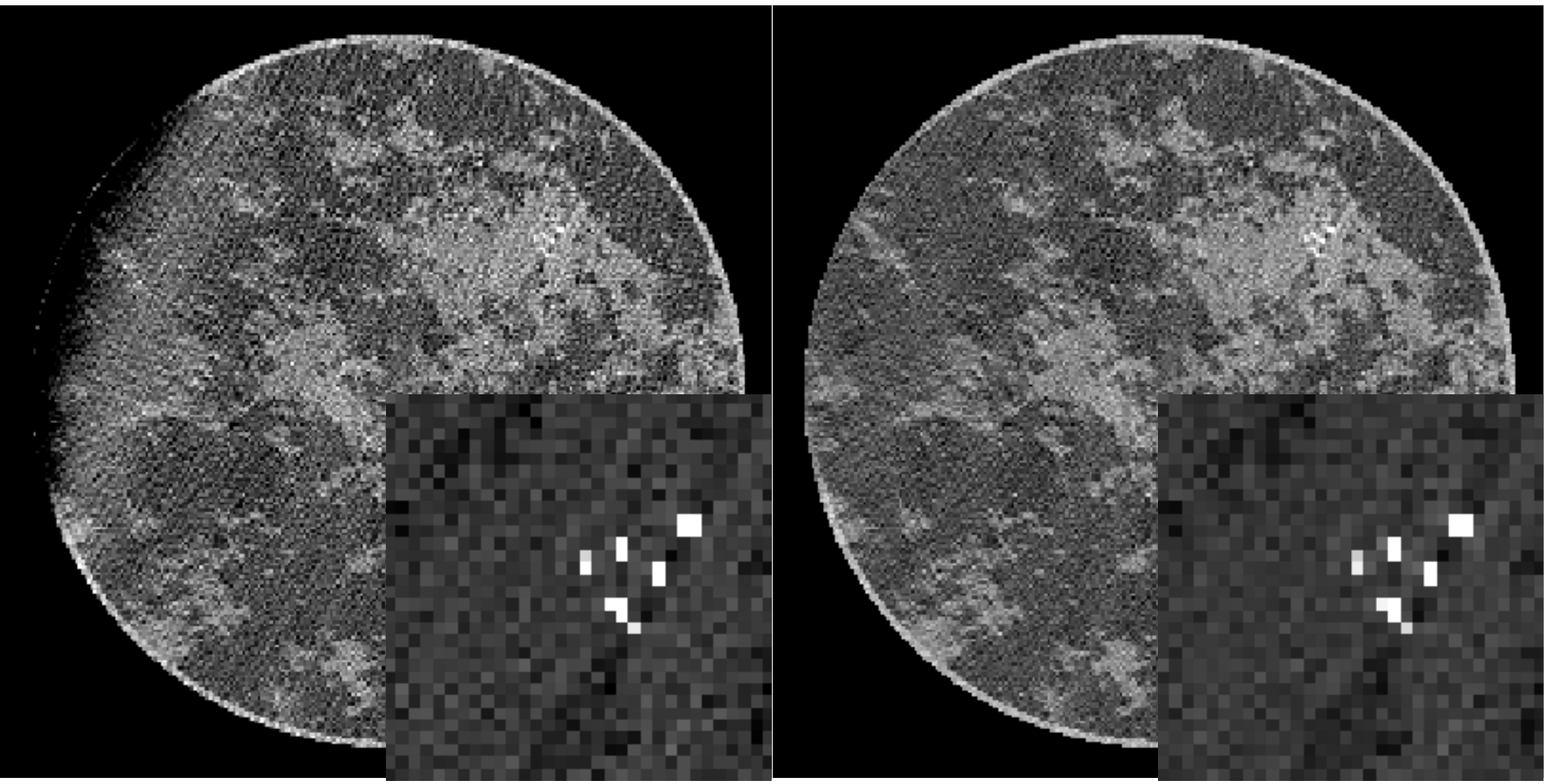}}
\end{minipage}
\caption{
Results of CP2-IC and CP1-IC
with noisy and inconsistent, simulated data. The curves labeled ``prior 0'' correspond
to a zero prior image. The curves labeled ``prior 1'' correspond to a prior image
of 1.0 on the object support. Top: (left) convergence of the data RMSE to the preset
value of $\epsilon=0.002$ and (right) image RMSE.
Bottom: (Left) ``prior 0'' final image, and (Right) ``prior 1'' final image.
Gray scales are the same as Fig. \ref{fig:phantom}.
The comparison between CP2-EC and CP1-EC
shows quantitatively the impact of the acceleration afforded by CP Algorithm 2.
\label{fig:noisyDataIneq}}
\end{figure}

The results of CP2-IC and CP1-IC are shown in Fig. \ref{fig:noisyDataIneq}.
The data RMSE is seen to converge
to the value established by the choice of $\epsilon^\prime$. In the displayed images, there is a clear
difference due to the choice of prior image. The image resulting from the zero prior
shows a substantial drift of the gray level on the left side of the image. Application of
a prior image consisting of constant background values over the object's true support removes
this artifact almost completely.
These results indicate that use of prior knowledge, when available,
can have a large impact
on image quality particularly for an ill-conditioned system matrix such as what arises in
limited angular-range CT.

\begin{figure}[!h]
\begin{minipage}[b]{0.45\linewidth}
\centering
\centerline{\includegraphics[width=\linewidth]{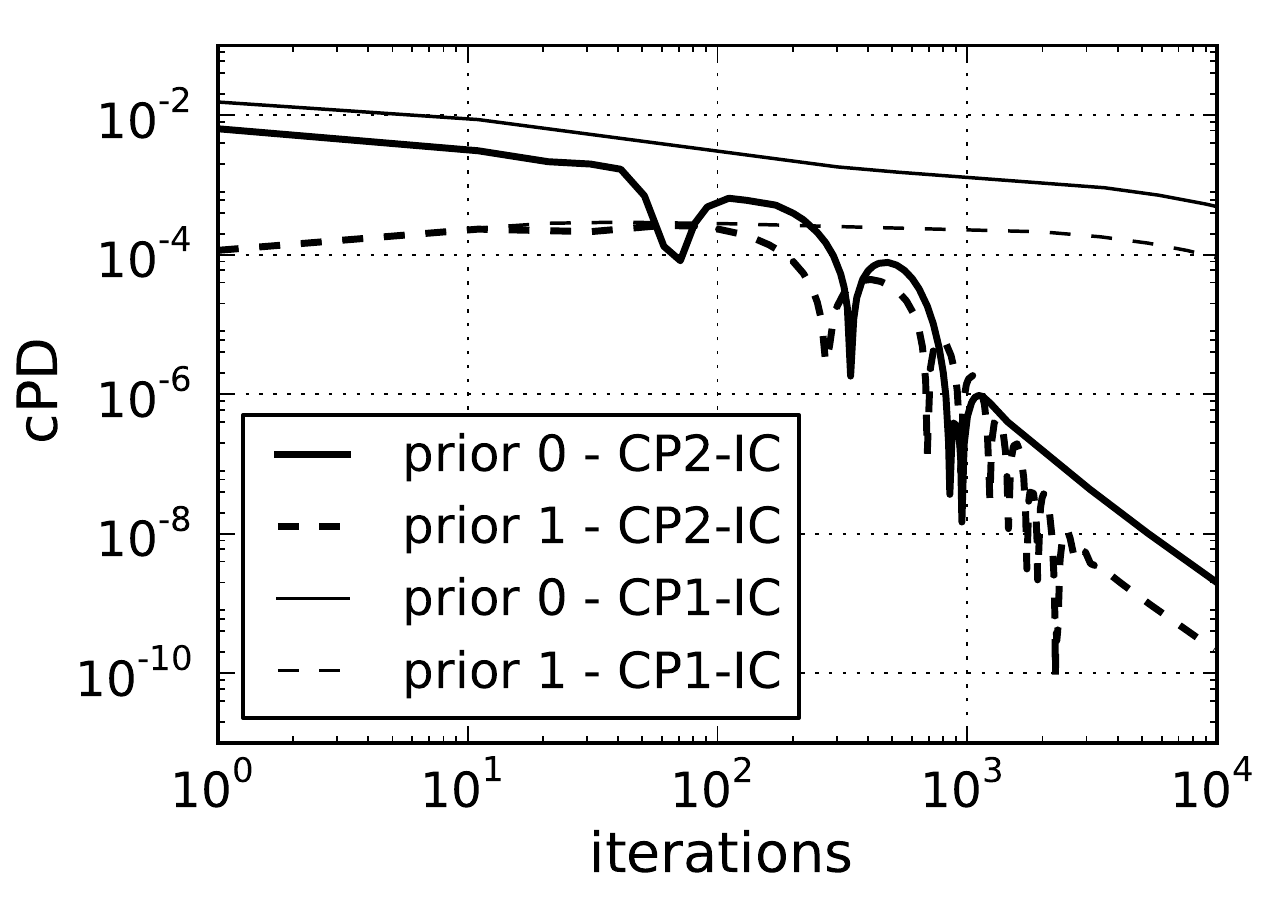}}
\end{minipage}
\caption{The conditional primal-dual gap for IC shown for CP2-IC and
CP1-IC. This gap is computed
by taking the difference between the primal and dual objective functions in Eqs. (\ref{cfxfgineq})
and (\ref{cfxfgineqdual}), respectively, after removing the indicator in the
primal objective function:
$cPD= \left|\frac{1}{2} \| \iv{f} - \iv{f}_\text{prior} \|^2_2 +
\frac{1}{2} \|X^T \iv{y} \|^2_2 + \epsilon^\prime \| \iv{y} \|_2
+\iv{g}^T \iv{y} - \iv{f}^T_\text{prior} (X^T \iv{y})
\right|/\text{size}(\iv{f})$. The absolute value is used because the argument can be negative,
and we normalize by the number of pixels $\text{size}(\iv{f})$ so that the primal objective function
takes the form of a mean square error. The prior image $\iv{f}_\text{prior}$  for this 
computation is explained in the text. The comparison between CP2-IC and CP1-IC
shows quantitatively the impact of the acceleration afforded by CP Algorithm 2.
\label{fig:ineqDataGap}}
\end{figure}

Because IC in this case presents a consistent problem,
convergence of the CP2-IC algorithm can be checked by the conditional primal-dual gap.
This convergence criterion is plotted for CP2-IC and CP1-IC in Fig. \ref{fig:ineqDataGap}.
The separate constraint check is seen in the data RMSE plot of Fig. \ref{fig:ineqDataGap}.
We see that the accelerated
version of the CP algorithm used in CP2-IC yields much more rapid convergence than CP1-IC.
For example, the data RMSE constraint is reached to within $10^{-6}$ at iteration 1000 for CP2-IC, while 
this point is not reached for CP1-IC by even iteration 10,000. A similar observation
can also be made for the conditional primal-dual gap.

\subsection{Noisy, inconsistent data with two-set convex feasibility}

For the last demonstration of the convex feasibility approach to IIR for limited-angular range
CT, we apply CP2-ICTV, which seeks
the image closest to a prior image and respects constraints on image TV and data-error.
We are unaware of other algorithms, which address this problem, and only results for CP2-ICTV
and CP1-ICTV are shown.
In applying CP2-ICTV, we need two constants, $\epsilon^\prime$ and $\gamma$, and accordingly use of this
algorithm is meant to be preceded by an initial image reconstruction in order to have a sense
of interesting values for the data-error and image TV constraints. From the previous results,
we already have information about data-error, and because we have the image estimates, we can
also compute image TV values.  The image TV values corresponding
to the two prior image estimates differ significantly, reflecting the quite different appearance
of the resulting images shown in Fig. \ref{fig:noisyDataIneq}. We follow the use of the support prior image,
and take the corresponding value of the image TV of 4,400.

\begin{figure}[!h]
\begin{minipage}[b]{0.45\linewidth}
\centering
\centerline{\includegraphics[width=\linewidth]{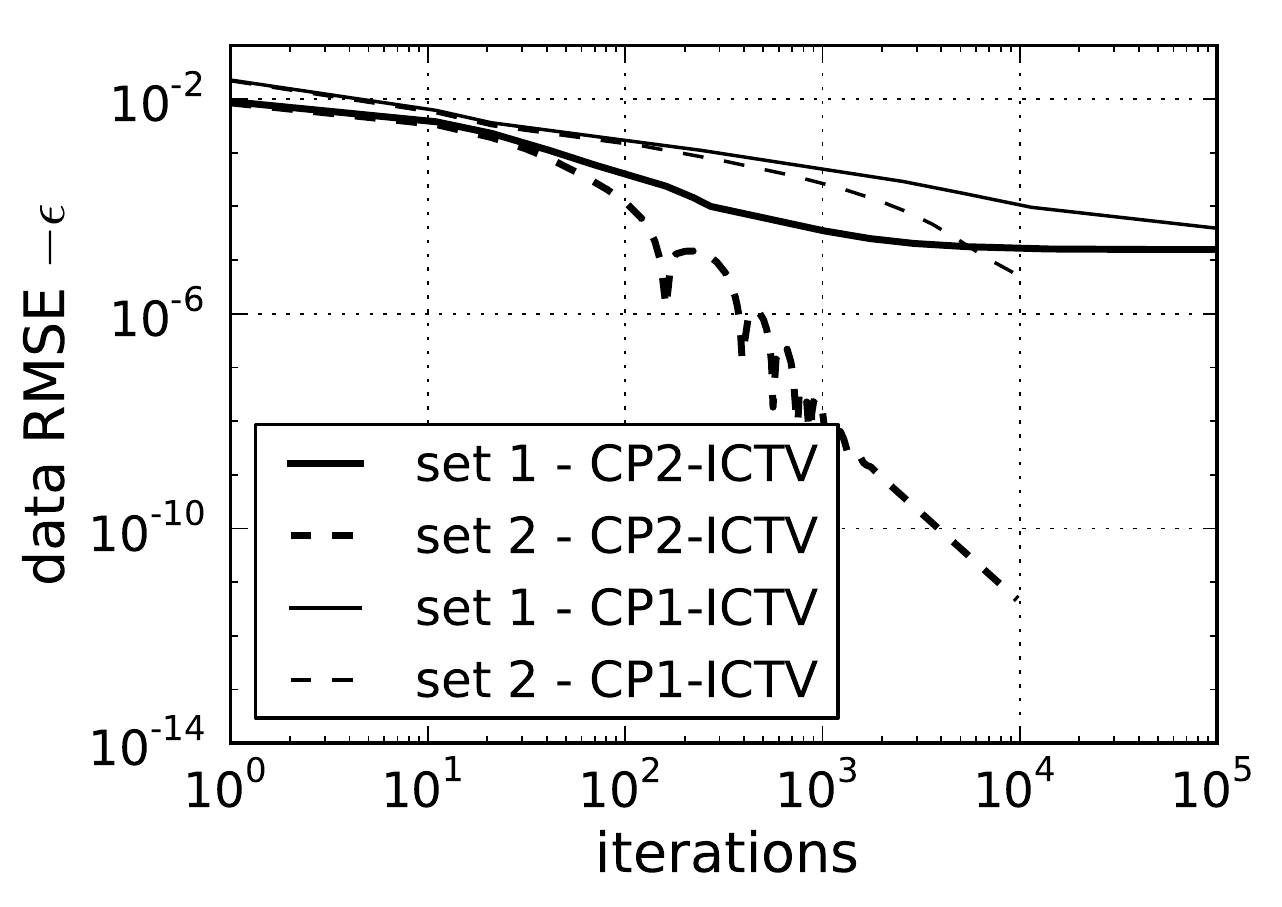}}
\end{minipage}
\begin{minipage}[b]{0.45\linewidth}
\centering
\centerline{\includegraphics[width=\linewidth]{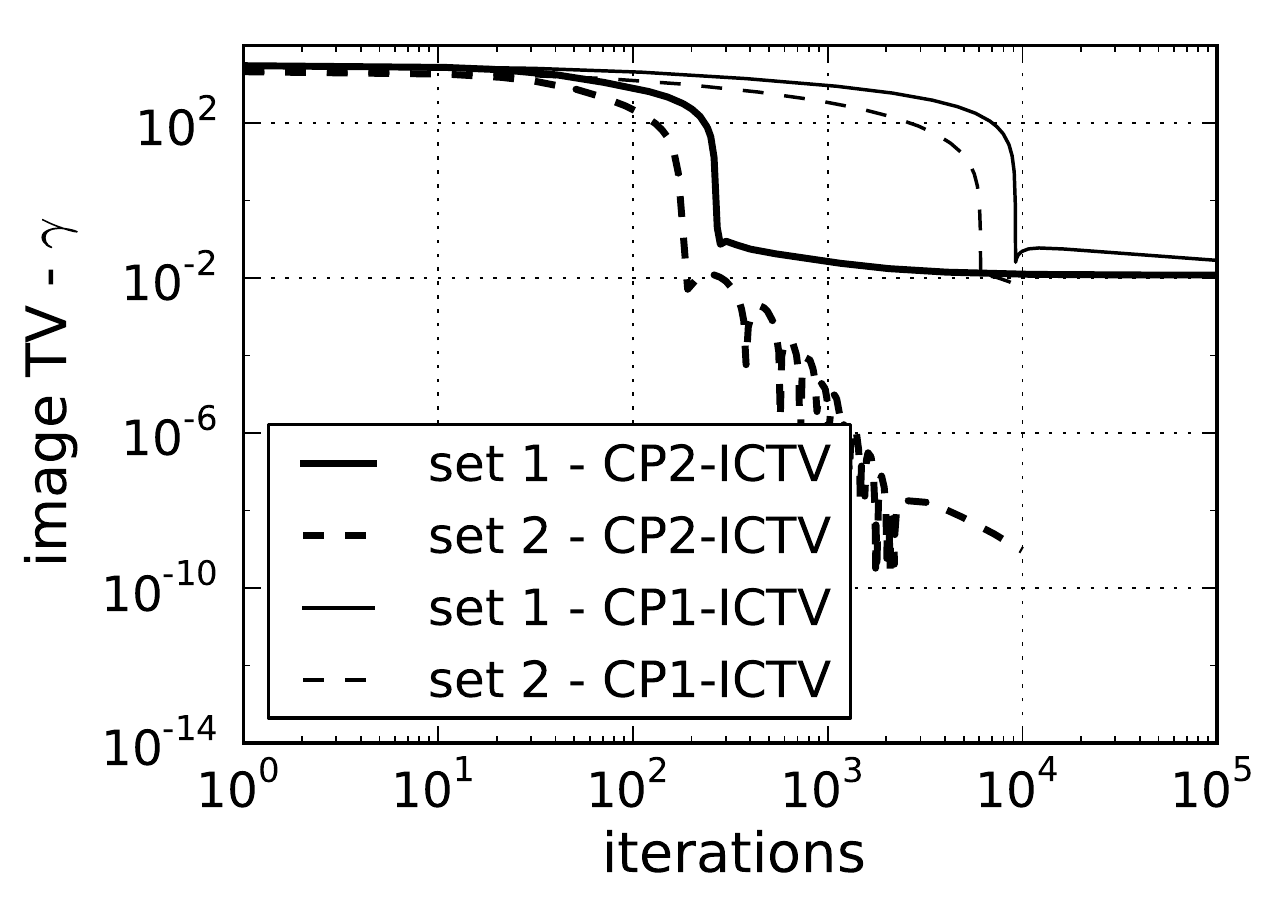}}
\end{minipage}
\begin{minipage}[b]{\linewidth}
\centering
\centerline{\includegraphics[width=0.45\linewidth]{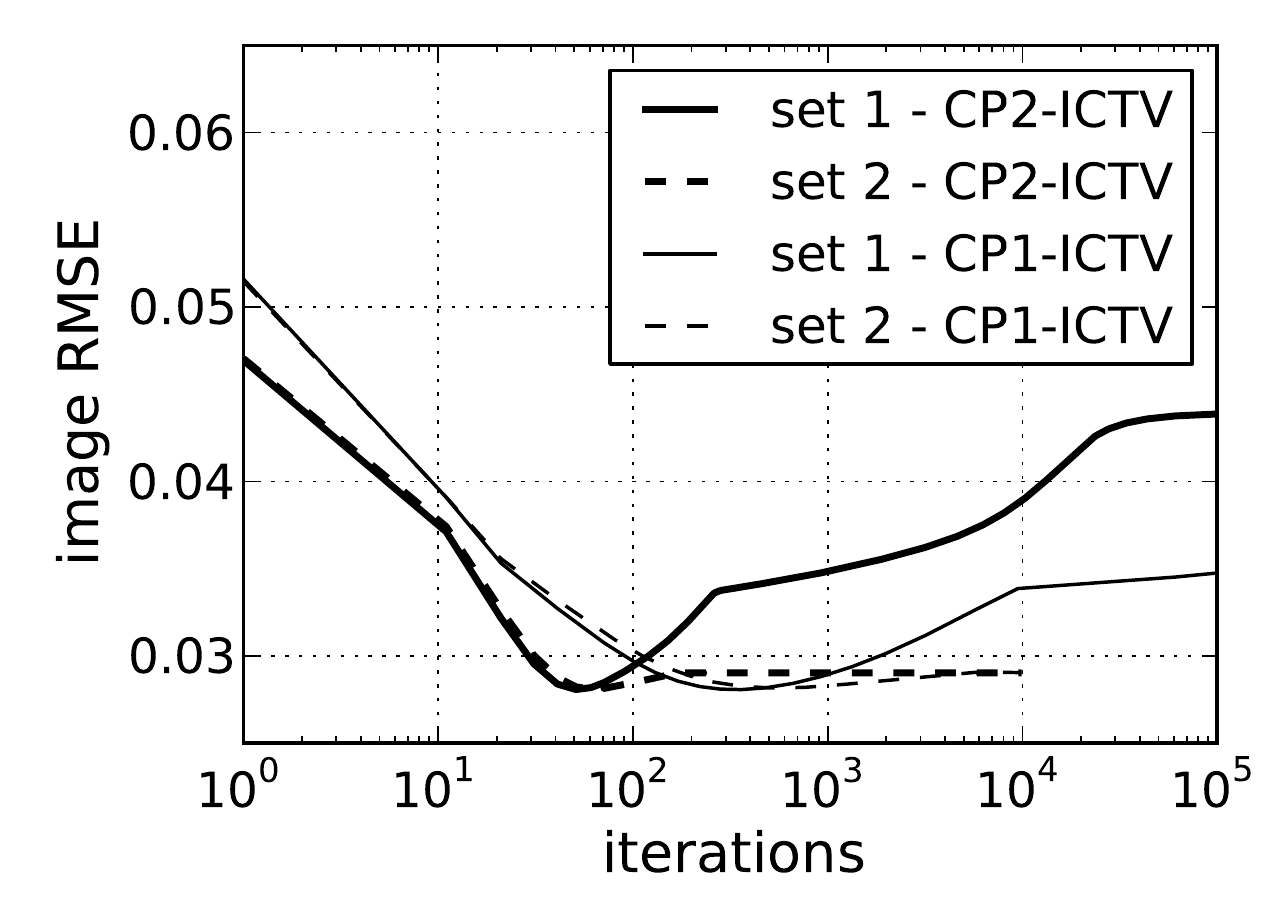}}
\end{minipage}
\begin{minipage}[b]{\linewidth}
\centering
\centerline{\includegraphics[width=0.6\linewidth]{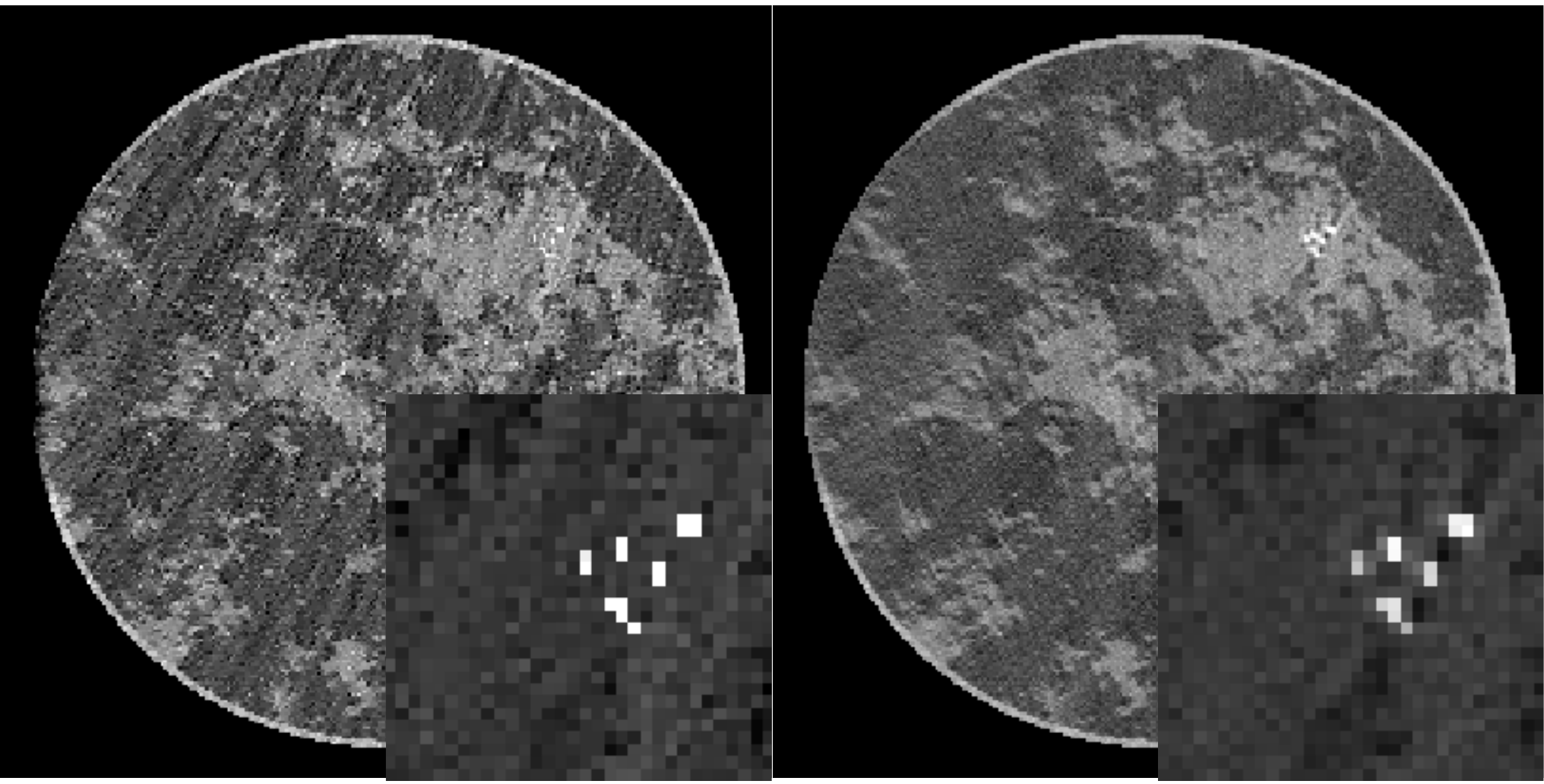}}
\end{minipage}
\caption{Results of CP2-ICTV and CP1-ICTV
with noisy and inconsistent, simulated data for
two different constraint set values: ``set 1'' refers to choosing $\epsilon^\prime =  0.512$
(a data RMSE of $\epsilon=0.002$) and $\gamma=4000$; ``set 2'' refers to choosing $\epsilon^\prime = 0.768$
(a data RMSE of $\epsilon=0.0025$) and $\gamma=3100$. Top row: (Left) evolution of data RMSE, and
(Right) evolution of image TV.
Middle row: evolution of image RMSE. The comparison between CP2-ICTV and CP1-ICTV
shows quantitatively the impact of the acceleration afforded by CP Algorithm 2.
Bottom row: (Left) resulting image of ``set 1'', and
(Right) resulting image of ``set 2''.
Gray scales are the same as Fig. \ref{fig:phantom}. Note that the calculation for ``set 1''
is extended to $10^5$ iterations due to slower convergence than the results for ``set 2.''
\label{fig:noisyDataTVxfg}}
\end{figure}

In our first example with this two-set convex feasibility problem, we maintain the tight
data-error constraint $\epsilon^\prime=0.512$ (a data RMSE of 0.002) but attempt to find an image
with lower TV by selecting $\gamma=4000$. The results for these constraint set
settings, labeled ``set 1'',
are shown in Fig. \ref{fig:noisyDataTVxfg}. Interestingly, this set of constraints appears to be
just barely infeasible; the CP2-ICTV
result converges to an image TV of 4000.012 and a data RMSE of 0.00202.
Furthermore, the dual variable magnitude increases steadily, an indication
of an infeasible problem. The curves for image TV and data RMSE indicate
convergence to the above-mentioned values, but we do not make theoretical
claims for 
convergence of the CP algorithms with inconsistent convex feasibility problems.


In the second example, we loosen the data-error constraint to $\epsilon^\prime=0.768$ (a data RMSE of 0.0025)
and seek an image with lower TV, $\gamma = 3100$, and the results are also shown in Fig. \ref{fig:noisyDataTVxfg}.
In this case, the constraint values are met by CP2-ICTV,
and the resulting image has noticeably less noise
than the images with no TV constraint imposed shown in Fig. \ref{fig:noisyDataIneq} particularly in the
ROI containing the model micro-calcifications. The image RMSE for this constraint set in ICTV is 0.029, while
the comparable image RMSE from the previous convex feasibility problem, IC, with no
TV constraint shown in Fig. \ref{fig:noisyDataIneq} is 0.037.
Thus we note a drop in image RMSE in adding the image TV constraint, but a true image quality
comparison would require parameter sweeps in $\epsilon$ for IC, and $\epsilon$ and $\gamma$ for ICTV.

\begin{figure}[!h]
\begin{minipage}[b]{0.45\linewidth}
\centering
\centerline{\includegraphics[width=\linewidth]{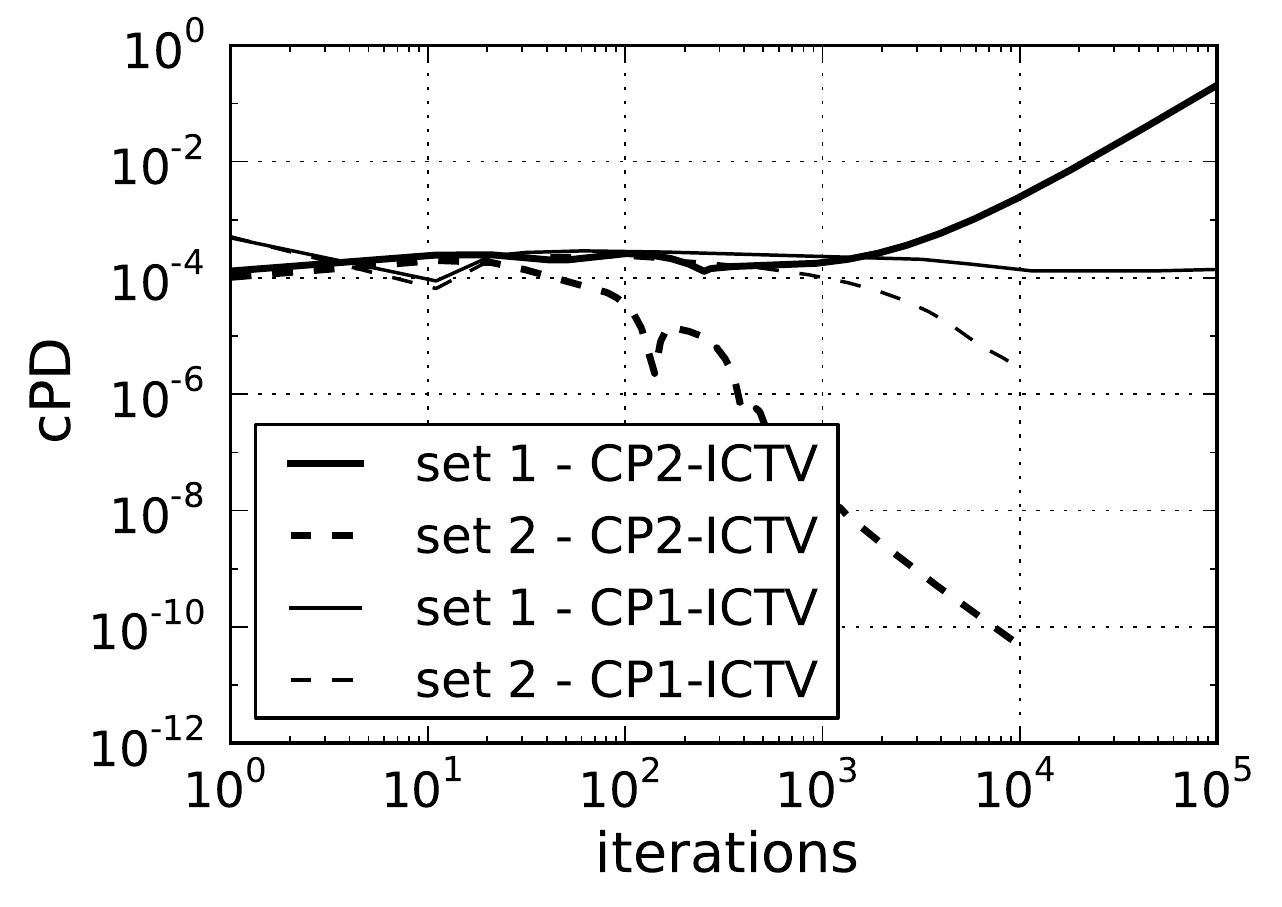}}
\end{minipage}
\caption{The conditional primal-dual gap for ICTV shown for CP2-ICTV and
CP Algorithm 1. This gap is computed
by taking the difference between the primal and dual objective functions in Eqs. (\ref{cfTVxfgineq})
and (\ref{cfTVxfgineqdual}), respectively, after removing the indicator in the
primal objective function:
$cPD= \left|\frac{1}{2} \| \iv{f} - \iv{f}_\text{prior} \|^2_2 +
\frac{1}{2} \|X^T \iv{y} \|^2_2 + \epsilon^\prime \| \iv{y} \|_2
+\gamma \| (|\iv{z}|) \|_{\infty}
+\iv{g}^T \iv{y} - \iv{f}^T_\text{prior} (X^T \iv{y}+ \nabla^T \iv{z})
\right|/\text{size}(\iv{f})$. The absolute value is used because the argument can be negative,
and we normalize by the number of pixels $\text{size}(\iv{f})$ so that the primal objective function
takes the form of a mean square error. The prior image $\iv{f}_\text{prior}$  for this 
computation is explained in the text. The comparison between CP2-ICTV and CP1-ICTV
shows quantitatively the impact of the acceleration afforded by CP Algorithm 2.
Note that the calculation for ``set 1''
is extended to $10^5$ iterations due to slower convergence than the results for ``set 2.''
\label{fig:TVineqDataGap}}
\end{figure}

Because this constraint set contains feasible solutions,
the conditional primal-dual gap can be used as a convergence check for CP2-ICTV.
This gap is shown for both sets of constraints in Fig. \ref{fig:TVineqDataGap}.
For CP2-ICTV there is a stark contrast in behavior between the two constraint sets. The feasible set
shows rapid convergence, while the infeasible set show no decay in the conditional primal-dual gap
below 1000 iterations and a steady increase from 1000 to 10,000 iterations. Again, the accelerated
CP algorithm used in CP2-ICTV yields a substantially faster convergence rate than CP1-ICTV for this
example.

\subsection{Comparison of algorithms}

With the previous simulations, we have illustrated use of the convex feasibility framework on EC, IC,
and ICTV for IIR in CT. The example for EC serves the purpose of demonstrating convergence properties of CP2-EC
on the ubiquitous least-squares minimization and establishing that this algorithm has competitive convergence rates
with standard algorithms, linear CG and ART. We do note that CG, on the shown example, does have the fastest
convergence rate, but the difference in convergence rate between CP2-EC, CG, and ART is substantially less
than their gap with the basic CP1-EC. 
For convex feasibility problems IC and ICTV, we have optimization problems where the current methodology can be
easily adapted to solve, but the standard algorithms linear CG and ART cannot easily be applied.
Because we have the comparisons of the CP algorithms on the EC simulations and because we have seen
convergence competitive with linear CG and ART, we speculate that CP2-IC and CP2-ICTV have competitive convergence
rates with any modification of CG or ART that could be applied to IC and ICTV.
In short, the convex feasibility framework using CP Algorithm 2 provides a means for proto-typing
a general class of optimization problems for IIR in CT, while having convergence rates competitive with
standard, but more narrowly applicable, large-scale solvers. Furthermore, concern over algorithm convergence
is particularly important for ill-conditioned system models such as those that arise in limited angular-range
CT scanning.


Convex feasibility presents a different design framework than unconstrained minimization or
mixed optimizations, combining e.g. data-fidelity objective functions with constraints.
For example, the field of compressed sensing (CS) \cite{candes2008introduction} has centered on devising
sparsity exploiting optimization for reduced sampling requirements in a host of imaging
applications. For CT, in particular, exploiting gradient magnitude
sparsity for IIR has garnered much attention, requiring the solution
to constrained, TV-minimization \cite{sidky2008image,SidkyCP:2012}
or TV-penalized, least-squares \cite{Jensen2011,Defrise:11,fessler:2012,SidkyCP:2012}.
The convex feasibility, ICTV, involves the same quantities but can be used only
indirectly for a CS-style optimization; the data-error can be fixed and multiple runs with
CP2-ICTV for different $\gamma$ can be performed with the goal of finding the minimum
$\gamma$ given the data and fixed-$\epsilon$. On the other hand, due to the fast convergence
of CP2-ICTV it may be possible to perform the necessary search over $\gamma$ faster
than use of an algorithm solving constrained, TV-minimization or a combined unconstrained objective function.
Also, use of ICTV provides direct control over the physical quantities in the optimization problem,
image TV and data-error, contrasting with the use of TV-penalized, least-squares, where
there is no clear connection between the smoothing parameter $\alpha$ and the final
image TV or data-error.
In summary, ICTV provides an alternative design for
TV-regularized IIR.

\section{Conclusion}
\label{sec:conclusion}

We have illustrated three examples of convex feasibility problems for IIR applied to limited
angular-range CT, which provide alternative designs to unconstrained or mixed optimization
problems formulated for IIR in CT.

One of the motivations of the alternative design is that
these convex feasibility problems are amenable to the accelerated CP algorithm, and
the resulting CP2-EC, CP2-IC, and CP2-ICTV algorithms solve their respective
convex feasibility problems with a favorable convergence rate-- an important feature for the ill-conditioned
data model corresponding the limited angular-range scan.  The competitive convergence rate is demonstrated
by comparing convergence of CP2-EC with known algorithms for large-scale optimization.
We then note that CP2-IC and CP2-ICTV, for which there is no alternative algorithm
that we know of, appears to have similar convergence rates to CP2-EC.

Aside
from the issue of convergence rate, algorithm design can benefit from the different
point of view offered by convex feasibility. For imaging applications this
design approach extends naturally to considering non-convex feasibility sets
\cite{luke2005relaxed,Han:2012}, which can have some advantage particularly for
very sparse data problems.  Future work will consider extension of the presented
methods to the non-convex case and application of the present methods to actual
data for CT acquired over a limited angular-range scan.

\begin{acknowledgments}
This work is part of the project CSI: Computational Science
in Imaging, supported by grant 274-07-0065 from the Danish
Research Council for Technology and Production Sciences.
This work was supported in part by NIH R01 grants CA158446, CA120540, and EB000225.
The contents of this article are
solely the responsibility of the authors and do not necessarily
represent the official views of the National Institutes of Health.
\end{acknowledgments}

\appendix

\section{Pseudocode for CP2-ICTV}
\label{sec:CF3-CP}

The pseudocode for CP2-ICTV appears in Fig. \ref{alg:cfTVxfgineq}, and 
we explain variables not appearing in Secs. \ref{sec:firstalg}
and \ref{sec:secondalg}. At Line \ref{auxstep} the symbol $\nabla$ represents
a numerical gradient computation, and it is a matrix which applies to an image
vector and yields a spatial-vector image, where the vector at each
pixel/voxel is either 2 or 3 dimensional depending on whether the image reconstruction
is being performed in 2 or 3 dimensions. Similarly, the variables $\iv{t}$ and
$\iv{z}_n$ are spatial-vector images. At Line \ref{dualupdate3B}
the operation ``$| \cdot |$'' computes the magnitude at each pixel
of a spatial-vector image, accepting a spatial-vector image and yielding
a scalar image. This operation is used, for example, to compute a gradient-magnitude image
from an image gradient. The ratio appearing inside the square brackets of Line \ref{dualupdate3B}
is to be understood as a pixel-wise division yielding a scalar image. It is possible
that at some pixels the numerator and denominator are both zero in which case we
define $0/0 = 1$. The quantity in the square brackets evaluates to a scalar image, which
then multiplies a spatial-vector image; this operation is carried out, again, in pixel-wise
fashion where the spatial-vector at each pixel of $\iv{t}$ is scaled by the corresponding pixel-value.
At Line \ref{primalupdate3}, $\nabla^T$ is the transpose of the matrix
$\nabla$, see Ref. \cite{SidkyCP:2012} for one possible implementation of $\nabla$
and $\nabla^T$ for two dimensions.

\begin{figure}
\hrulefill
\begin{algorithmic}[1]
\State $L \gets \| (\sm{X},\nabla) \|_2; \; \tau \gets 1; \; \sigma \gets 1/L^2; \; n \gets 0$
\State initialize $\iv{f}_0$,  $\iv{y}_0$, and $\iv{z}_0$ to zero vectors
\State $\bar{\iv{f}}_0 \gets \iv{f}_0$
\Repeat
\State $\iv{y}^\prime_n \gets  \iv{y}_n+\sigma( \sm{X} \bar{\iv{f}}_n -\iv{g})$; 
$ \; \iv{y}_{n+1} \gets \max( \| \iv{y}^\prime_n\|_2 - \sigma \epsilon^\prime, 0 )
\frac{\iv{y}^\prime_n}{ \| \iv{y}^\prime_n \|_2} $ \label{dualupdate3A}
\State $\iv{t} \gets \iv{z}_n +\sigma \nabla \iv{f}_n$ \label{auxstep}
\State $\iv{z}_{n+1} \gets \iv{t} \,
\left[ \left( |\iv{t}| - \sigma \, \text{proj}_{\text{Diamond}(\gamma)} (|\iv{t}|/\sigma) \right)
/|\iv{t}| \right]$
\label{dualupdate3B}
\State $\iv{f}_{n+1} \gets  \left[ \iv{f}_n - \tau (\sm{X}^T \iv{y}_{n+1} - \iv{f}_\text{prior} +
\nabla^T \iv{z}_{n+1}) \right] / (1+\tau)$ \label{primalupdate3}
\State $ \theta \gets 1/\sqrt{1+2\tau}$; $\tau \gets \tau \theta$; $\sigma \gets \sigma / \theta$ 
\State $\bar{\iv{f}}_{n+1} \gets \iv{f}_{n+1} + \theta(\iv{f}_{n+1} - \iv{f}_n)$
\State $n \gets n+1$
\Until{$n \ge N$}
\end{algorithmic}
\hrulefill
\caption{Pseudocode for $N$ steps of the accelerated CP algorithm instance
for solving Eq. (\ref{cfTVxfgineq}) with parameters $\epsilon^\prime$ and $\gamma$.
Variables are explained in the text, and pseudocode for the function $\text{proj}_{\text{Diamond}(\gamma)}(\iv{x})$ is
given in Fig. \ref{alg:projell1}.}
\label{alg:cfTVxfgineq}
\end{figure}

The pseudocode for the function $\text{proj}_{\text{Diamond}(\gamma)} (\iv{x})$ appears in
Fig. \ref{alg:projell1}. This function is essentially the same as
what is listed in Figure 1 of Ref. \cite{duchi2008efficient}; we include it here
for completeness. The ``if'' statement at Line 2, checks if the input vector $\iv{x}$
is already in $\text{Diamond}(\gamma)$. Also, because the function $\text{proj}_{\text{Diamond}(\gamma)} (\iv{x})$ is used
with a non-negative vector argument in Line \ref{dualupdate3B} of Fig. \ref{alg:cfTVxfgineq},
the multiplication by $\text{sign}(\iv{x})$ at the end of the algorithm
in Fig. \ref{alg:projell1} is unnecessary for the
present application. But we include this $\text{sign}(\iv{x})$ factor so that the function applies to any
$N$-dimensional vector.

\begin{figure}
\hrulefill
\begin{algorithmic}[1]
\Function{$\text{proj}_{\text{Diamond}(\gamma)}$}{$\iv{x}$}
\If {$\|\iv{x}\|_1 \le \gamma$}
   \State \Return{$\iv{x}$}
\EndIf
\State $\iv{m} = |\iv{x}|$
\State Sort $\iv{m}$ in descending order: $m_1 \ge m_2 \ge \dots m_N$
\State $\rho \gets \max j$ such that $m_j - \frac{1}{j}
\left( \sum_{k=1}^{j} m_k - \gamma \right) >0$, for $j \in [1,N]$
\State $\theta \gets (1/\rho) \left( \sum_{k=1}^{\rho} m_k -\gamma \right)$
\State $\iv{w} = \max ( |\iv{x}| - \theta, 0)$
\State \Return{$\iv{w} \, \text{sign}(\iv{x})$}
\EndFunction
\end{algorithmic}
\hrulefill
\caption{Pseudocode for the function $\text{proj}_{\text{Diamond}(\gamma)} (\iv{x})$, which projects
$\iv{x}$ onto the $\ell_1$-ball of scale $\gamma$. This function appears at line
\ref{dualupdate3B} of algorithm in Fig. \ref{alg:cfTVxfgineq}. The vector $\iv{x}$ is taken
to be one-dimensional with length $N$, and the individual components are labeled $x_i$
with index $i$ being an integer in the interval $[1,N]$.}
\label{alg:projell1}
\end{figure}

\newpage

\bibliographystyle{Medphys}
\bibliography{convexProto}

\end{document}